\documentclass{article}
\usepackage{arxiv}
\usepackage[utf8]{inputenc} % allow utf-8 input
\usepackage[T1]{fontenc}    % use 8-bit T1 fonts
\usepackage{hyperref}       % hyperlinks
\usepackage{url}            % simple URL typesetting
\usepackage{booktabs}       % professional-quality tables      
\usepackage{amsmath,amssymb,amsfonts}
\usepackage{algorithmic}              % blackboard math
\usepackage{nicefrac}       % compact symbols for 1/2, etc.
\usepackage{microtype}      % microtypography
\usepackage{lipsum}
\usepackage{graphicx}
\graphicspath{ {./images/} }
\usepackage{lscape}
\usepackage{subfigure}

\title{Energy and Time Based Topology Control Approach to Enhance the Lifetime of WSN in an economic zone}

\author{
Tanvir Hossain\textsuperscript{1}, 
Md. Ershadul Haque\textsuperscript{2}, 
Abdullah Al Mamun\textsuperscript{1}, 
Samiul Ul Hoque\textsuperscript{1}, 
Al Amin Fahim\textsuperscript{1}
\\
\textsuperscript{1}Department of Electrical \& Electronic Engineering, Feni University, Feni-3900, Bangladesh.\\
Email: tanvirhossainzillu@gmail.com, mamun.eee.fu@gmail.com, samiul0906@gmail.com,\\
mdalaminfahim61@gmail.com
\\
\textsuperscript{2}School of Computing, Mathematics and Engineering, Charles Sturt University, Bathurst, Australia.\\
Email: mhaque@csu.edu.au
}

\begin{document}
\maketitle

\begin{abstract}

An economic zone requires continuous monitoring and controlling by an autonomous surveillance system for heightening its production competency and security. Wireless sensor network (WSN) has swiftly grown popularity over the world for uninterruptedly monitoring and controlling a system. Sensor devices, the main elements of WSN, are given limited amount of energy, which leads the network to limited lifespan. Therefore, the most significant challenge is to increase the lifespan of a WSN system. Topology control mechanism (TCM) is a renowned method to enhance the lifespan of WSN. This paper proposes an approach to extend the lifetime of WSN for an economic area, targeting an economic zone in Bangladesh. Observations are made on the performance of the network lifetime considering the individual combinations of the TCM protocols and comparative investigation between the time and energy triggering strategy of TCM protocols. Results reveal the network makes a better performance in the case of A3 protocol while using the topology maintenance protocols with both time and energy triggering methods. Moreover, the performance of the A3 and DGETRec is superior to the other combinations of TCM protocols. Hence, the WSN system can be able to serve better connectivity coverage in the target economic zone.

\end{abstract}

\keywords{Wireless Sensor Network, Topology Control Algorithm, Economic Zone, Wireless Communication.}

\section{Introduction}
\label{Introduction}
In the productive infrastructures, i.e, economic zone, industry, garment, farming fields, etc. the measure and taking care of different types of physical quantities, such as, humidity, temperature, pH level, stress, vibration and so on are the most important factors that escalate the proficiency of the control system in that target area. Although traditional wired communication techniques have long been used to support these circumstance, wireless communication techniques are now being widely utilized due to its convenient implementation, low installation and maintenance cost, etc. Recent advancements in technologies have been possible to develop autonomous, energy-efficient and intelligent wireless sensor nodes that can be distributed in large numbers in a geographical territory to make a self-oriented and self-healing network scenario called wireless sensor network (WSN).

Sensor nodes, the main element of WSN, in general are assumed to have three major ability, namely, sensing, computation, and wireless communication. The role of sensing capability is to acquire the data of physical quantities from the environment. On the other hand, necessary data aggregation, control information processing, and managing of sensing and communication activities are conducted in the computational unit. Wireless communication methods, finally, transmit as well as receive data and control information from one sensor node to the other node. A sensor node, at a particular time, remains involved with operating one of finite set probable activities or asleep. Every task executed by a sensor nodes costs a definite amount of energy, and the nodes in sleeping mode do not execute any task and consume, of course, no amount of energy. However, a sensor node performing in the WSN system subject to several fundamental curbs as follow \cite{R1}, 
\begin{itemize}
    \item Each sensor node is not aware of its time identification of fabrication.
    \item Sensor nodes are made so that their size can be minimized as possible and produced in mass amount where testing is a complex procedure.
    \item Every sensor nodes given a fixed amount of energy equivalent to its built in power supply expired after its power being exhausted.
    \item Each sensor node has internal timer settings that control it to remain asleep most of the time to save energy except for some random instants when it wakes up with functionality.
    \item The sensor node has considerable transmission range, generally, in meters, that covers a few sensor nodes deployed in the network.
    \item It is not feasible or practical to make attention to individual sensor nodes while they are deployed.
\end{itemize}

The traditional interest in using WSN has focused on high-end applications, i.e. nuclear radiation leakage surveillance, seismic and climate data monitoring, armed and medical applications, etc. But, the most recent aspects widely involve the WSN systems to be utilized in the applications extended from the consumers levels to the national and international security systems. The applications of WSN in existence and potentially being developed include mainly physical security, pollution protection, agriculture monitoring, traffics management, industrial, production and business automation, smart home and city, electric power plant and substations, infrastructure monitoring, etc. \cite{R2,R3}.

The main contributions including this approach cover the following points:
\begin{itemize}
    \item To design a WSN system for monitoring and controlling an economic region based on autonomous and real-time facilities.
    \item to find out an optimized solution for the node deployment, with the WSN system making minimum energy consumption and prolonging its lifetime.
    \item To evaluate and compare the performance of the WSN system considering its two maintenance criteria, such as time and energy.
\end{itemize}

The remains of this paper are as: \textit{Literature Review} section summarizes most related works from the part of the literature. Another section, \textit{The Basic Architecture of WSN System}, provides an overview of the WSN architecture. \textit{Mathematical Model} characterizes the equations and formulas of various models used in WSN system, i.e., communication model, energy dissipation model, etc. The section, \textit{Methodology}, provides an insight into the principles, methods and techniques of this work. The findings and discussions are presented in the section called \textit{Performance of the WSN}. \textit{Conclusion} includes a concise description of the work and a few future directions.

\section{Literature review}
\label{Literature}

There are a lot of research works found in the literature on WSN based communication technology, revealing the recent trends, parameters affecting, application specific domains, constraints, benefits, etc.
\\
The authors introduce a hybrid WSN using two types of sensor nodes, i.e. static and mobile nodes for a mid size urban city. A web server hosted external database manages all of the gateways to be synchronized. Actually mobile nodes carries out two major responsibility like data transmitting and concentrating that validate the placement of the gateway across the deployment. Temperature, humidity, pressure, gases, global positioning system (GPS) tracking, etc. are allowed to be real time monitored with its associated location in the city \cite{R4}.
\\
In \cite{R5}, a method by WSN to find out the probability of fire as soon as possible before it is happened in a forest is proposed. Through this WSN founded detection process, deploying a number of sensor nodes in the forest whereabouts, forest authority can easily access the remote location of the forest in view to monitoring the climate conditions. These nodes are categorized into two difference, for example, GPS integrated and not GPS integrated to make the model cost efficient. High active (HA), medium Active (MA), and low active (LA) zones split a forest into three diverse area. In order to reserve the energy consumption of the network, HA, continuously and MA, in periodic, communicate with the base station but LA does not link up with the base station. The system accuracy results in 90\% while sensing one parameter by the nodes. 
\\
A watching that the framers have to transcend a colossal loss for imprecise weather prediction and imbalance nursing of the crops and vegetables is often in the field. A self acting agricultural ambiances surveillance, controlling the relevant factors in the ground, by WSN may reduce the cost of production and raise the sprouting. Expecting to promote yield, a report on WSN based tomato cultivation was presented in paper \cite{R6}. An automotive irrigation system facilitates the water management discipline including the control of the water flowing, real time feedback of water level in the storage, etc. At a 5 seconds interval, the desired data from the farmer comfortable monitoring of agricultural circumstance become updated to a web server, how a visitor can look after the growing from anywhere.    
\\
In view to monitoring the industrial wastewater needed to be discharged, keeping track of potential of hydrogen (pH) value, oxygen quantity, conductivity, etc., authors demonstrate a cloud unified WSN in \cite{R7}. It gathers the sensed data in ThingSpeak, a cloud service, through global system for mobile communications (GSM)/general packet radio service (GPRS) networking to conduct real time investigation and integration to internet of things (IoT). Telerivet messaging, a short message service (SMS) notification circulating framework, alerts the occurrence of water pollution to the respective bodies, appending more benefits to the system.  
\\
Countries undergoing rapid industrialization and urbanization are often witnessing an increase in air pollution. Due to the salvation of hazardous gases, such as carbon monoxide ($CO$), carbon dioxide ($CO_2$) and sulfur dioxide ($SO_2$), especially from the industrial establishments, both human health and the environment remain under intimidation over the years. The aim at \cite{R8} is to give down a cost and energy effective WSN for looking after the qualities of air in the environment. The architectures of various components of the system, for example, sensor node, microcontroller, wireless module, software tool, etc. are detailed in the article. ThingSpeak, too, utilized in this work carries out the real time data processing to the users. Hierarchical Based Genetic Algorithm (HBGA) set to have an efficacy of the energy, one of the major Challenges for WSN after the constraint energy of the sensor.    
\\
Article \cite{R9} enunciates a WSN applied noise pollution metering technique's development proceedings and a noise hygiene monitoring process. Recently noise pollution concerning vehicles, airplanes, construction apparatus, factories, electrical devices, etc. has emerged at a menacing level, affecting millions around the world. Sensor's findings can be summed in a central database center in order to get real time access to the information required through a graphical user interface. JTS 13-57 digital sound noise measuring meter including ATmega328 micro controller, GPS, 1.5 Km transmission capable XBEE are devoted to the architecture of WSN. 
\\
Description of a WSN's development for observing the healthcare's most pivotal factors is dictated by \cite{R10}. It incorporates a centralized node and four sensor nodes, with well instrumented sensing devices, i.e. Humidity, temperature, $CO$, $CO_2$, infra-red, etc. sensors. Four sensor nodes and a centralized node need to be equipped in the four patient's rooms and the nursing base, respectively in a hospital site. The system is seen to have a root mean squared error of 1.1\%, $0.35^o$C, 0.98\% in case of humidity, temperature, and gas, respectively.
\\
The authors provide the article \cite{R11} with discussing an intelligent method using WSN system for localizing and estimating boundary of the gas leakage sources in an underground coal quarry. Poisonous and inflammable gases in the coalmine, for instance, CO, Methane ($CH_4$), not only damaging the health but leading to an explosion, rehash personnels' life threatening. The concentration magnitude of the area pointed gases is analyzed considering Gaussian Plume Gas Model (GPGM) to identify the source of leakage, boundary, and gass' congelation grid map. 
\\
\cite{R12} hosts a representation of the WSN architecture for smart grid that covers the power's uninterrupted, automated, and reliable management. Power quality and voltage rising problems was controlled using a dynamic controller. Data required after being measured on the feeders is provided to a database center through continuously run web server.  
\\
WSN can be regarded to be one of the reliable and real time communication systems for military applications, having to oversee a large remote territory including forces, ammunition, protective weapons, vehicles, etc. \cite{R13} leaves a WSN operated military surveillance system to secure the militia bases lying in danger. Nodes' random deployment and physical inaccessibility are the two most challenging issues in this case. Hence, the energy effective functions of the WSN are optimized by an exact node scheduling method.

\section{The basic architecture of WSN System}
\label{Architecture }
\begin{figure}[h]
\centering
\includegraphics[width=0.35\linewidth]{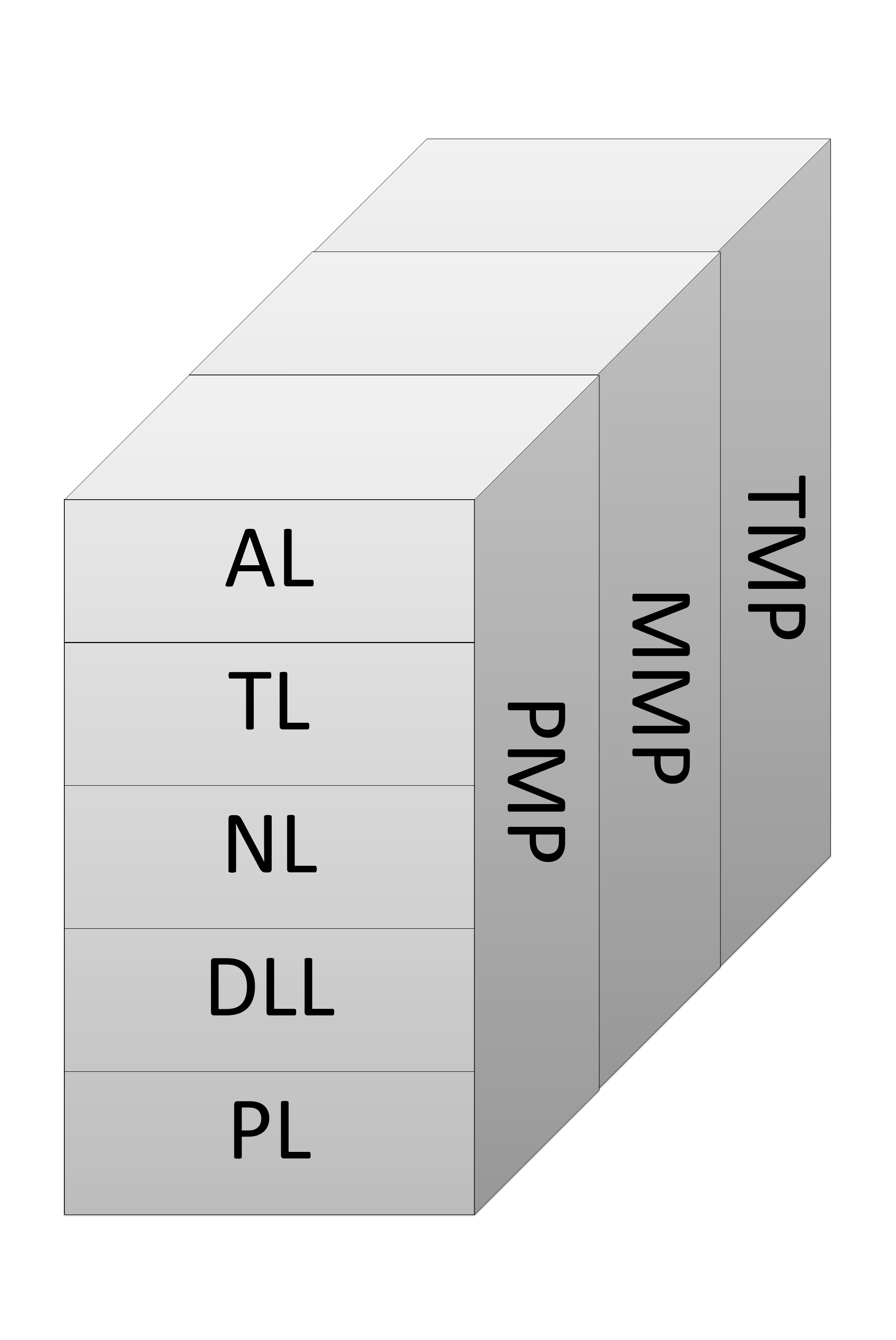}
\caption{OSI/LCA architecture of WSN}
\label{OSI_LCA_architecture}    
\end{figure}
The Open Systems Interconnection (OSI) model, also familiar as Layered Communication Architecture (LCA), is the most popular architecture adopted by WSN \cite{R14}. This architecture involves both sink node and source node to take part into the transmission of data packets.
\\
Materially, OSI/LCA is constructed of five layers, i.e. Application Layer (AL), Transport Layer (TL), Network Layer (NL), Data Link Layer (DLL), and Physical Layer (PL). Furthermore, three cross-plane layers, such as, Power Management Plane (PMP), Mobility Management Plane (MMP), and Task Management Plane (TMP) are also integrated in conjunction with the above mentioned five layers \cite{R15,R16}. It is made sense better the concept of a typical WSN architecture in the figure \ref{OSI_LCA_architecture}.
\\
PL: It is utilized to produce the interface while transmitting the data packets through a physical channel. The major tasks of PL are selecting, detecting, and generating frequency. Additionally, electrical and mechanical interfaces are also described by this layer. IEEE 802.15.4 standard is intended in regard to the layer to prolong the lifetime of the batteries considering lower energy consumption and cost, simplicity, and communication range etc. for a WSN system \cite{R17}.
\\
DLL: This layer makes sure a dependable communication between the point-to-point connection and the point-to-multipoint connections. It also conducts the data multiplexing operation, the medium access control, detecting a frame, and the controlling of errors. In pursuance of the application kind, several tasks, such as, optimization of throughput, efficiently energy consumption as possible, self-maintaining through hop-to-hop communication etc. are too performed by the layer.
\\
NL: Basically the NL is organized based on a few principal concepts, i.e. energy consumption, data aggregation, data centric characteristics of WSN, location of the nodes etc. It uses different types of routing protocols based on minimum energy, minimum hop, maximum available energy etc. in order to economize the energy consumption. Another significant job done by the NL is it joins the network with external connections. 
\\
TL: As the implementation of a WSN need to be considered according to the nature of applications, so TL layer is to be designed accordingly. The main task of TL is to meet the necessity of reliability and avoidance of congestion. Some protocols applied on not only upstream but also downstream are developed using the mechanism for loss detection and recovery. Due to the limitation of energy and memory of the nodes, the designing of TL is raised to a challenge \cite{R18}. 
\\
AL: AL is responsible for the time-synchronizing and traffic-managing operation, configuring the nodes (i.e. information about nodes), security, and so on \cite{R19}. The layer also draws management-level activities depending upon the need of application and user.
\\
PMP is used to deal with the energy of the nodes when it continues to consume during the operation of the WSN. The job of MMP is to discover the mobility of the nodes and control the neighbor nodes. TMP is given to make the arrangements for the sensing events within a certain area \cite{R20,R21}.

\section{Mathematical Model}
\label{mathmetical_model}
\begin{figure}[h]
\centering
\includegraphics[width=0.80\linewidth]{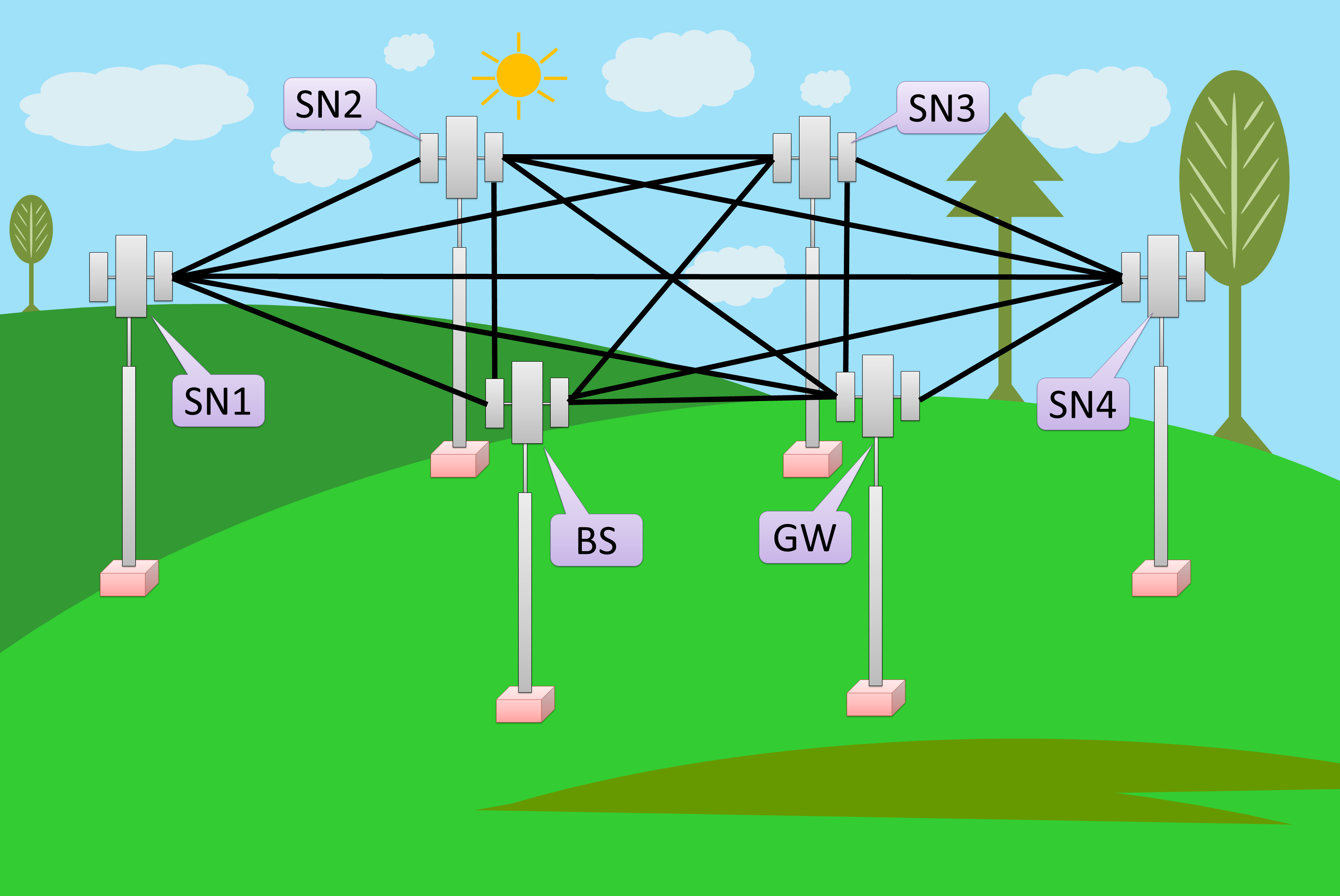}
\caption{Dense network scenario of the topology.}
\label{Dense_network}    
\end{figure}
\subsection{Communication model}
Consider a signal arrived at a receiver after the transmitter had transmitted it, traveling a distance $d$. At the receiver end, the power of the signal can be defined by \eqref{equation_path_loss_model}, according to the Two Ray Ground Propagation Model \cite{R22} which yields a precise result \cite{R23}. Here, transmitter and receiver antennas' heights are respectively $h_t$ and $h_r$ and the gains of those corresponds to $G_t$ and $G_r$, respectively. $P_t$ is the power of the transmitted signal at the end of the transmitter. $C_t$, a constant, depends upon the sensitivity of the transceiver. Letting the equation \eqref{equation_path_loss_model} interpret that the coverage of the transmitter can be regarded as a circle consist of diameter $d$ where, 
\begin{equation}
P_r = \frac {P_t.G_t.G_r.{h_t}^2.{h_r}^2} {d^4} = C_t \frac{P_t}{d^4}
\label{equation_path_loss_model}
\end{equation}
\begin{equation}
d = 2\sqrt[4]{C_t.P_t}
\end{equation}
\subsection{Energy dissipation model}
The energy decayed in the transmitter and receiver device can be estimated by the First Order Radio Model \cite{R24}, a mostly used in the general energy dissipation model. Consider a transmitter, with amplifiers and associated electronic circuits, consumes $E_{tx}$ energy which can be determined by equation \ref{Energy_dissipation_model_transmitter} while transmitting $k$ bit data at $l$ distance. Noted that transmitter circuit's and amplifier's used energy is $E_{ele-tx}(k)$ and $E_{amp-tx}(k,l)$, respectively. The receiver remaining on the other side draws the energy given by the equation \ref{Energy_dissipation_model_receiver}. The receiver device only care for the electronic circuit, of which energy expenditure $E{ele-rx}(k)$ is akin to the spent energy of the receiver's electronic circuit, according to the model.
\begin{equation}
E_{tx} = E_{ele-tx}(k)+E_{amp-tx}(k,l)
E_{tx} = E_{ele}.k+E_{amp}.k.l^2
\label{Energy_dissipation_model_transmitter}
\end{equation}
\begin{equation}
E_{rx} = E_{ele-rx}(k)
E_{rx} = E_{ele}.k
\label{Energy_dissipation_model_receiver}
\end{equation}
\subsection{Sensing model}
Let an circular area be divided into three circular zone, i.e. interior, intermediate, and outer zone. Three zones are assumed to have a different probability of sensing, basing on the range of the sensor. In conformity with probabilistic sensing model, the interior zone and the outer zone have a probability tends to one and zero, respectively and the probability in the intermediate zone exponentially decays with respect to distance \cite{R25}. It comes out in mathematical formation with equation \ref{Sensing_coverage}. Where, $C(s)$ is sensing coverage of an area like a disk of radius $r$. $\alpha=x-(r-r_u)$ and $x$ is the euclidean distance from sensor to an event. $\lambda$ and $\beta$ rely on the type of physical sensor devices.
\begin{equation}
C(s) =
\begin{cases}
1,\ for \ r-r_u \geq x \\
e^{-\lambda \alpha^{\beta}},\ for \ r-r_u<x \leq r+r_u\\
0,\ for \ r+r_u<x
\end{cases}
\label{Sensing_coverage}
\end{equation}
\subsection{Critical transmission range}
Critical transmission range (CTR) is one of the most well known methods of controlling the transmitting signal's power centrally. CTR takes into account minimum communication range of all nodes in a network. Suppose that $n$ nodes are allotted to a square edged area of length $l$. CTR under a two dimensional dense network can be estimated by the Penrose formula, in equation \eqref{CTR} \cite{R26}. Given $f(n)$ is a dependent function to $n$ such that $\lim_{n \to \infty} f(n)= +\infty$ and $log(n)$ is natural logarithm of $n(ln(n))$.
\begin{equation}
CTR_{dense} = \sqrt{\frac{log(n)+f(n)}{n\pi}}
\label{CTR}
\end{equation}

\section{Methodology}
\label{methodology}
\begin{figure}[h]
\centering
\includegraphics[width=0.60\linewidth]{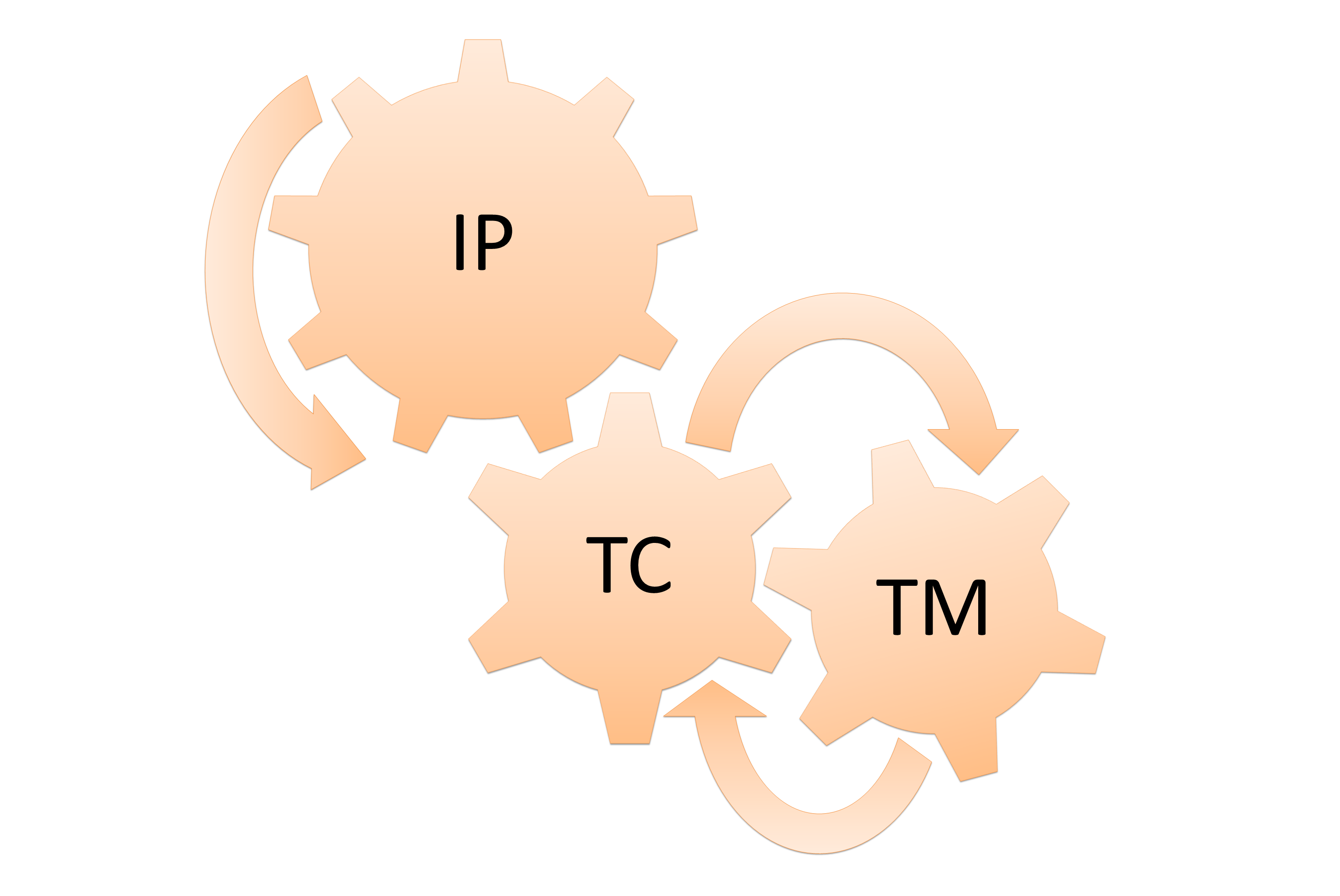}
\caption{A schematic representation of TCM cycle}
\label{TCM_cycle}    
\end{figure}
\begin{figure}[h]
\centering
\includegraphics[width=0.35\linewidth]{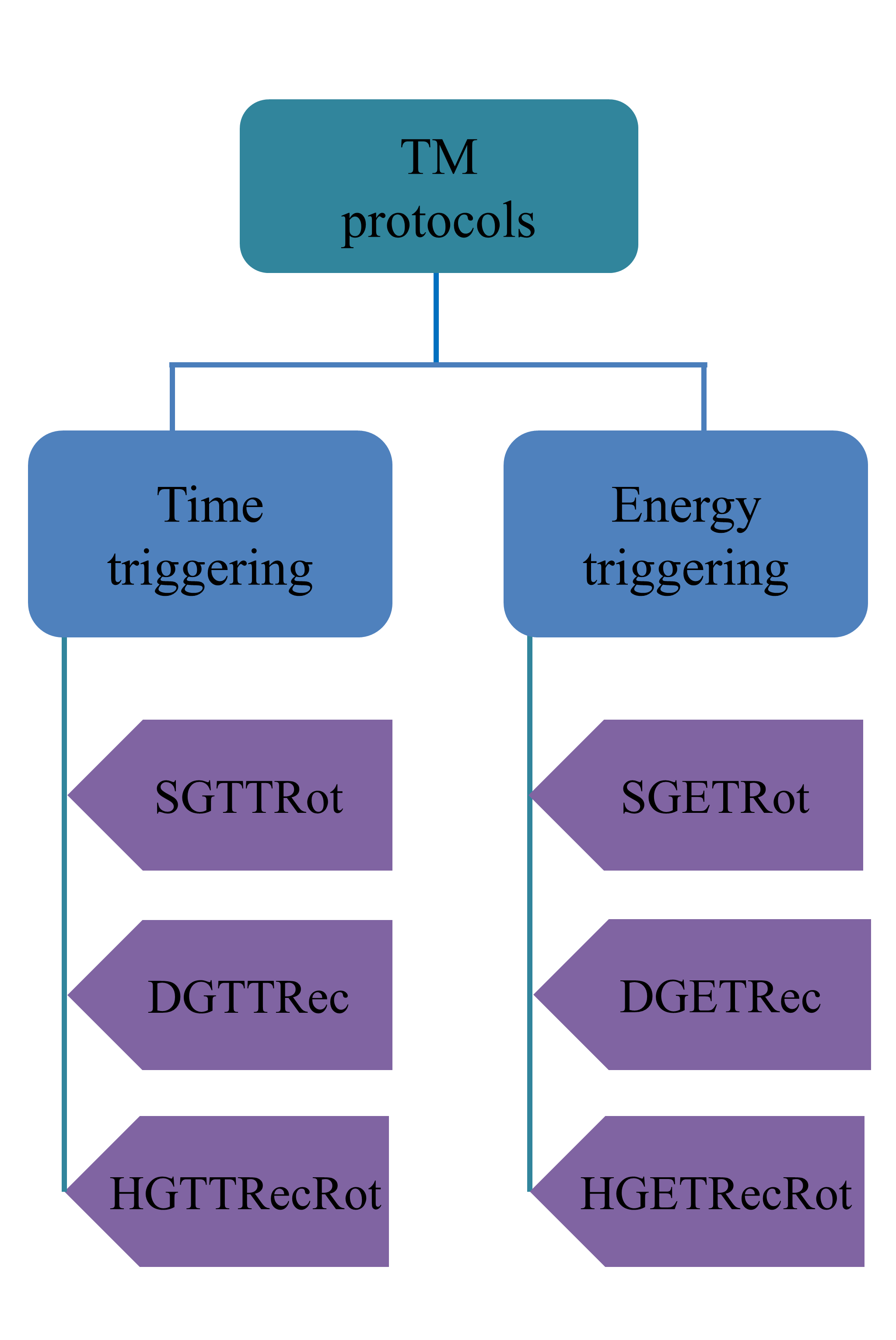}
\caption{TM protocols, categorized into two orders.}
\label{Categorized_TM_protocols}    
\end{figure}

To lengthen the lifetime of a WSN is still a challenging task as it is an application specific communication system. Even though different types of WSN algorithm, such as, topology construction (TC) algorithms, topology maintenance (TM) algorithms are developed, their exact infliction, either combined or separately, is one of the most crucial catalysts for a certain area of application. The WSN system using the topology control mechanism (TCM) is designed under the dense network scenario, which is led by TC and TM algorithms.
\\
The dense deployment of the nodes dictates that each of the nodes has a group of neighbors taken part in sensing the same event. Figure \ref{Dense_network} demonstrates an instance of the dense network to have six nodes deployed in a region. It reveals that all of the nodes in the network are directly connected to the base station. Furthermore they have also directly interconnection among themselves.
\\
TCM is generally defined to be a process that has an iterative progress as given by the figure \ref{TCM_cycle}. Firstly an initialization phase (IP) similar to all WSN deployment is performed. The operation of IP is like that all the nodes under the deployment detect themselves to find the initial topology by making use of their maximum transmission power. After the IP, the TC the next phase, a newly discovered reduced topology (RT) is built. Since the energy of the sensor nodes participated to the network will be consumed over the time, RT will never be longer active after a confined time. No sooner had the TC phase established RT than the activity should be started by TM phase next to TC phase.  At that time a protocol must be needed to observe the status of the RT and round the TC phase at the required time. TCM expects the cycle to be repeated in view to enhancing the lifetime of a network as long as it has energy.
\begin{table}[h]
\caption{TCM protocols}
\centering
%\begin{tabular}[htbp]{@{}ll@{}}
\begin{tabular}[h]{@{}ll@{}}
\hline
\textbf{Parameters} & \textbf{Design values}\\
\hline
TC protocols & A3, A3Cov\\
TM protocols & DGETRec, HGETRecRot, SGETRot\\
 &  DGTTRec, HGTTRecRot, SGTTRot\\
Routine protocols & Simple forwarding\\
Aggregation protocols & Simple S and D\\
\hline
\end{tabular}
\label{TCM_protocols}
\end{table}
In this work, to be TC protocol, A3 and A3cov algorithms are used who using communication coverage and sensing coverage, respectively of the nodes. On the other hand, TM protocols are applied, categorizing into two classes based on the triggering criteria, such as energy and time as shown in figure \ref{Categorized_TM_protocols}. TCM protocols, on the whole, including additional protocols too are given in the table \ref{TCM_protocols}.

\begin{figure}[h]
\centering
\includegraphics[width=0.80\linewidth]{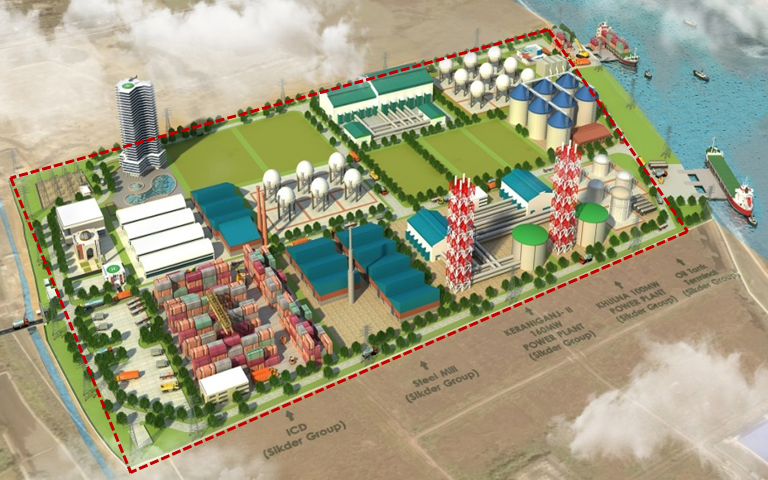}
\caption{Site selection: PowerPac Economic Zone}
\label{Site_selection}    
\end{figure}

PowerPac Economic Zone, an economic region, dwelt in Bangladesh is contemplated for the proposed approach \cite{R27}. The area holding length 1074m and width 660m of the economic zone is considered as checked by the red dotted line in figure \ref{Site_selection}. 300 nodes are spread out over the target domain on the basis of uniform distribution model conferring a locus on the sink node at the center. To sum up, the characteristics of the site are encapsulated in table \ref{Features_of_deployment}.

\begin{table}[h]
\caption{Features of the deployment area.}
\centering
%\begin{tabular}[htbp]{@{}lc@{}}
\begin{tabular}[h]{@{}lc@{}}
\hline
\textbf{Parameters} & \textbf{Design values}\\
\hline
Number of nodes & 300\\
Area of deployment & $1074m\times660m$\\
Distribution model & Uniform\\
Sink location & Center\\
\hline
\end{tabular}
\label{Features_of_deployment}
\end{table}

\begin{figure*}[h]
\centering
\includegraphics[width=0.80\linewidth]{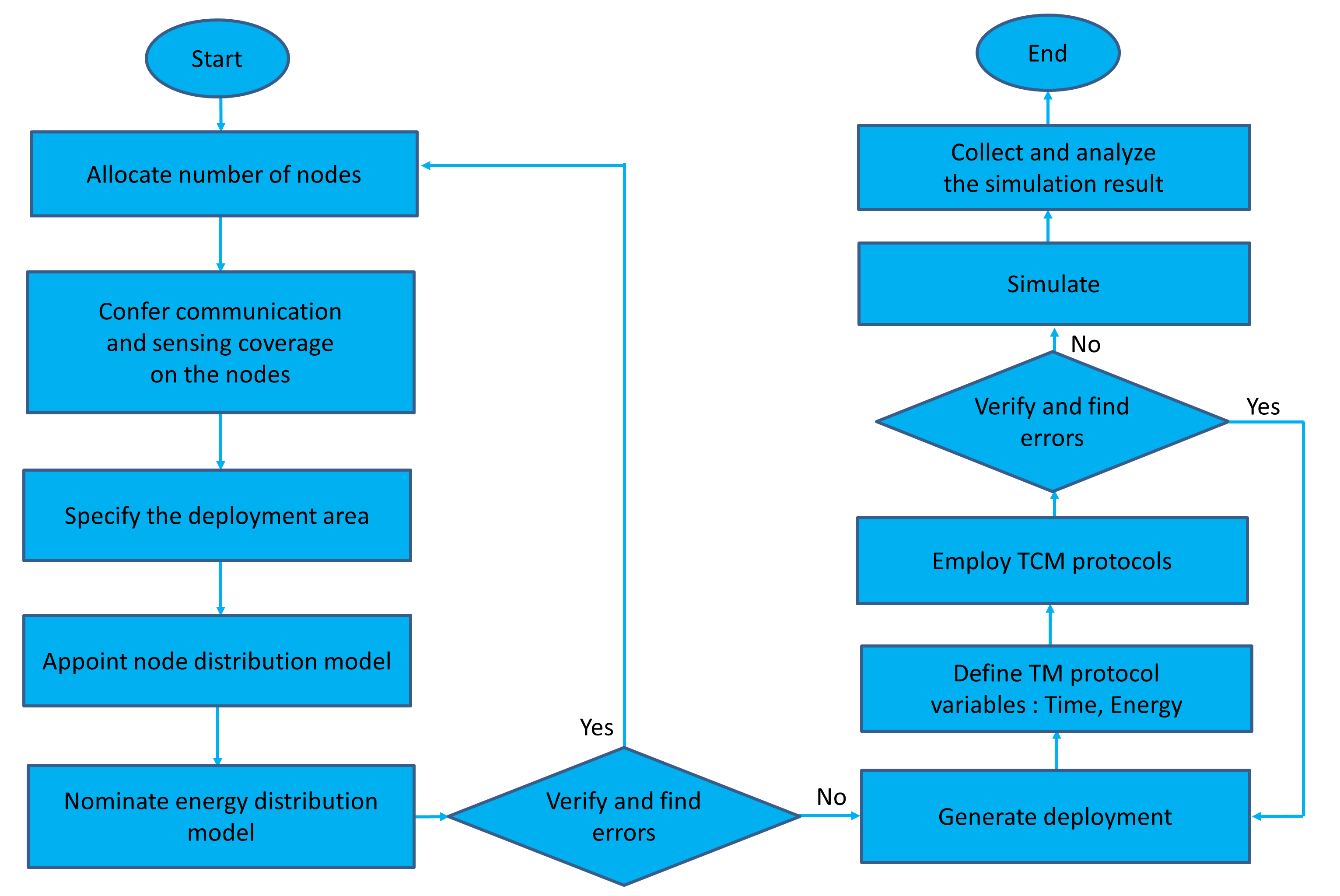}
\caption{The flowchart while conveying the an easy elucidation of the methodology.} 
\label{flow_chart}    
\end{figure*}

Based on the transmitting power and the energy consumption of the nodes, the communication and the sensing radius of a node is defined 100m and 20m, respectively. Internal energy outlay results from the EC and the AC, according to the node energy model. Therefore, energy of the EC and AC is decided upon $50nJ/bit$ and $10pJ/bit/m^2$. Howover, it is summarized in the table \ref{Design_values_of_nodes}.
\\
\begin{table}[h]
\caption{Design values of the nodes}
\centering
%\begin{tabular}[htbp]{@{}ll@{}}
\begin{tabular}[h]{@{}ll@{}}
\hline
\textbf{Parameters} & \textbf{Design value}\\
\hline
Communication radius & 100m\\
Sensing radius & 20m\\
Initial Energy Source & $E_{ele}=50nJ/bit$\\
                      & $E_{amp}=10pJ/bit/m^2$\\
\hline
\end{tabular}
\label{Design_values_of_nodes}
\end{table}
The way the approach proceeds is illustrated in the figure \ref{flow_chart}. In the initiation, an adequate number of nodes requires to be allocated which would be deployed. After that, the area for the deployment and the transmission power of the nodes as well as their sensing range are defined. Then, allocating the node and energy distribution model, an investigation is conducted to find whether there are any errors or not. In case no errors, the deployment is generated, otherwise the steps need to iterate. The TM variables, such as time and energy threshold are ascertained in the next step. Another surveillance is placed after the TCM protocols being conferred. If no errors are pointed out, simulation is driven. At the end, the performances of the network come by are analyzed.

\section{Performances of the WSN}
\label{result}
The major incentive for TCM is to edify the lifetime of a WSN system that is able to economize the consumption of energy, reserving the most beneficial peculiarities of the networks, such as coverage, sensitivity, etc.
\\
The nodes deployed in a two-dimensional region have a communication disk without any flaws. The information related to the position, orientation, and neighbor of the nodes are unknown to each other. An initial connected graph is constructed right after the deployment takes shape. A metric proportional to the received signal strength indicator (RSSI) can be utilized to enumerate the distance. DLL exists with having no packet losses. Mechanisms are adjoined to awaken a node amid its radio turned off. These assumptions are to be made certain while evaluating the performances.
\\
Following performance metrics are taken into discourse, such as,
\begin{itemize}
    \item Utilization of nodes.
    \item Active nodes linked to the sink.
    \item Connectivity of the area.
    \item Sensitivity of the area.
\end{itemize}
Basically, inspections are rendered upon the two criteria of the TM protocols based on the TC protocols, as given,  
 
\subsection{Energy triggering criteria}
\subsubsection{A3 protocol}
\begin{figure*}[h]
    \centering
    \subfigure[]
    {
        \includegraphics[width=0.45\linewidth]{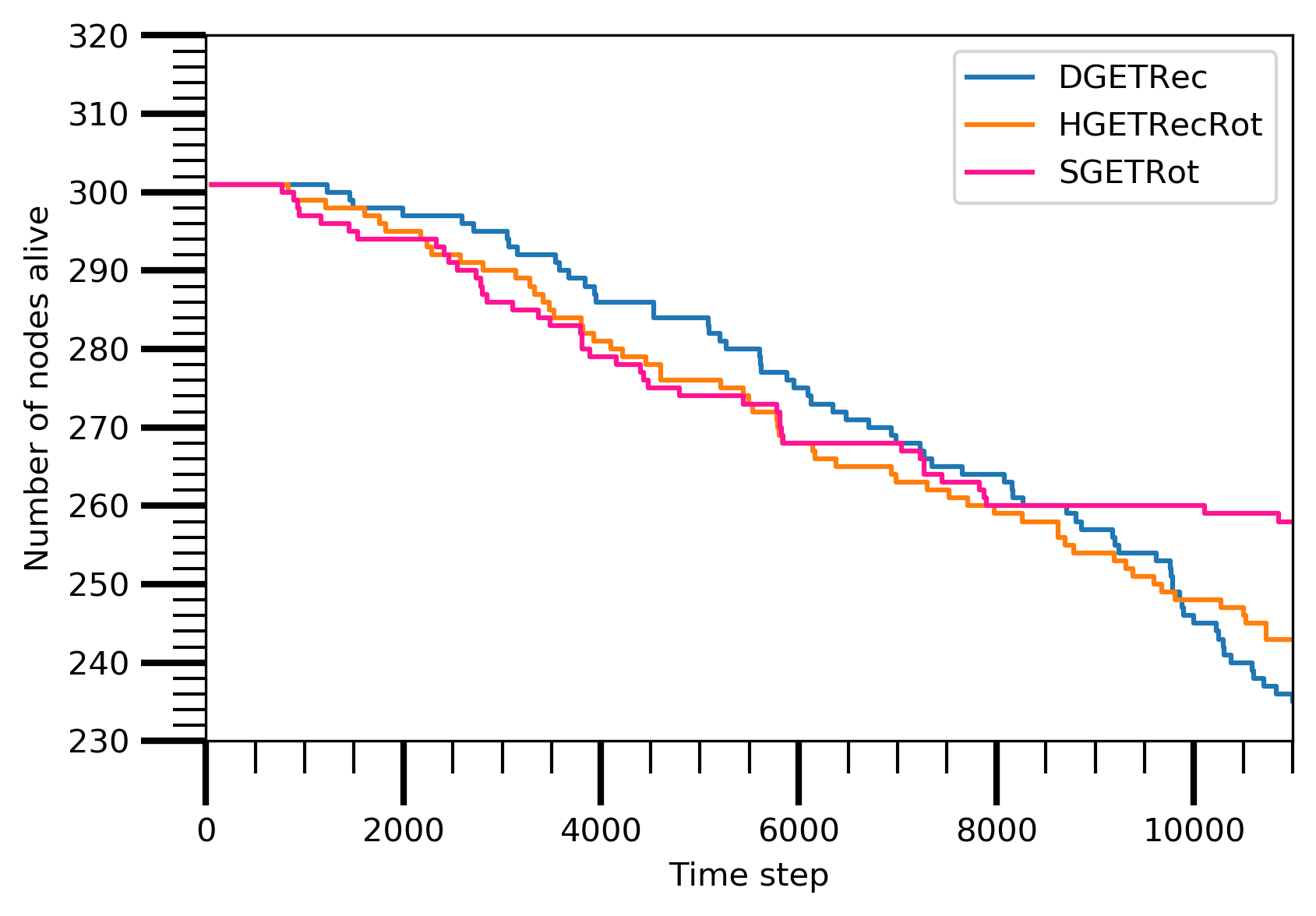}
        \label{A3_energy_alive_nodes}
    }
    \subfigure[]
    {
        \includegraphics[width=0.45\linewidth]{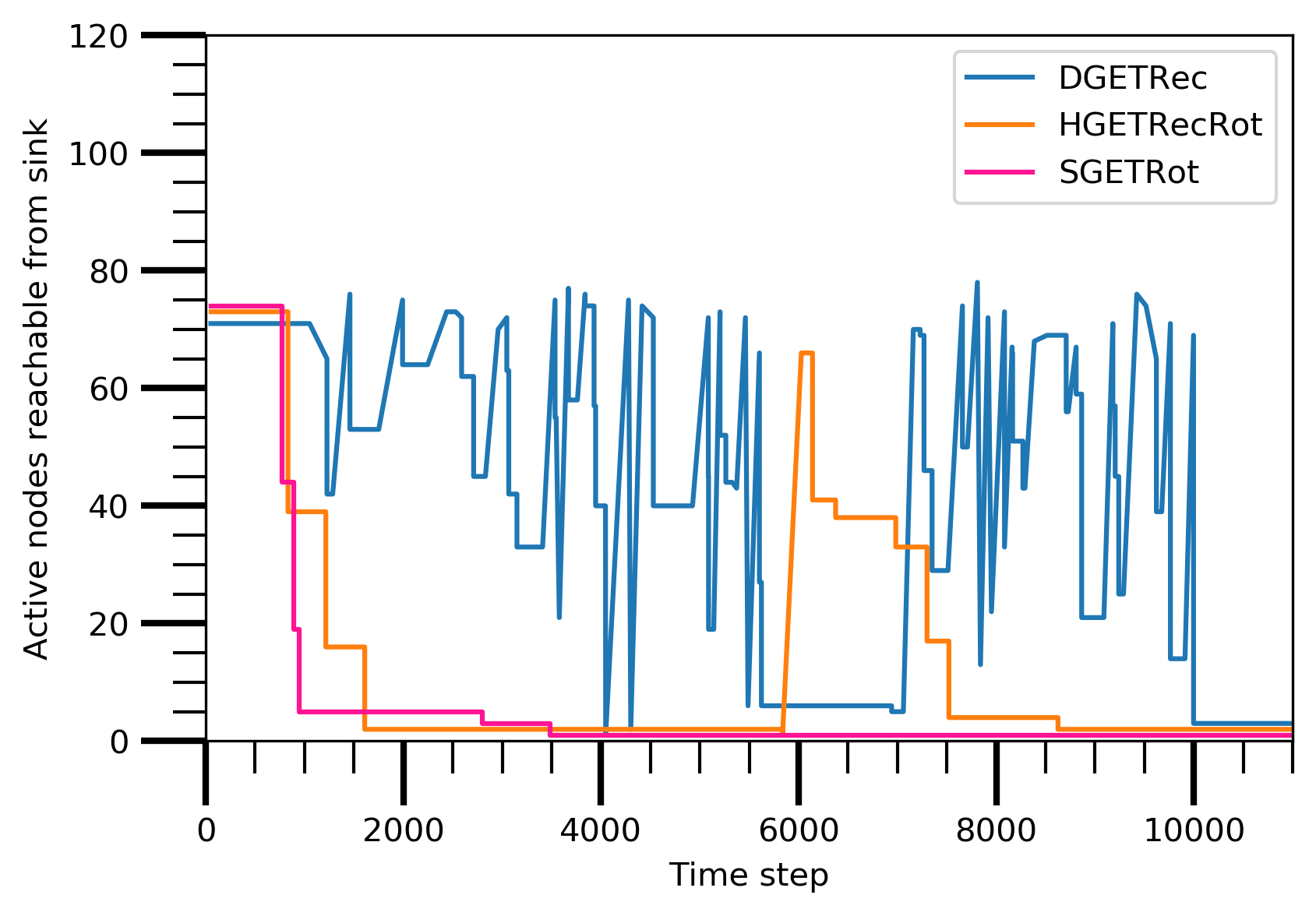}
        \label{A3_energy_Number_of_node_reachable_from_sink}
    }
    \\
     \subfigure[]
    {
        \includegraphics[width=0.45\linewidth]{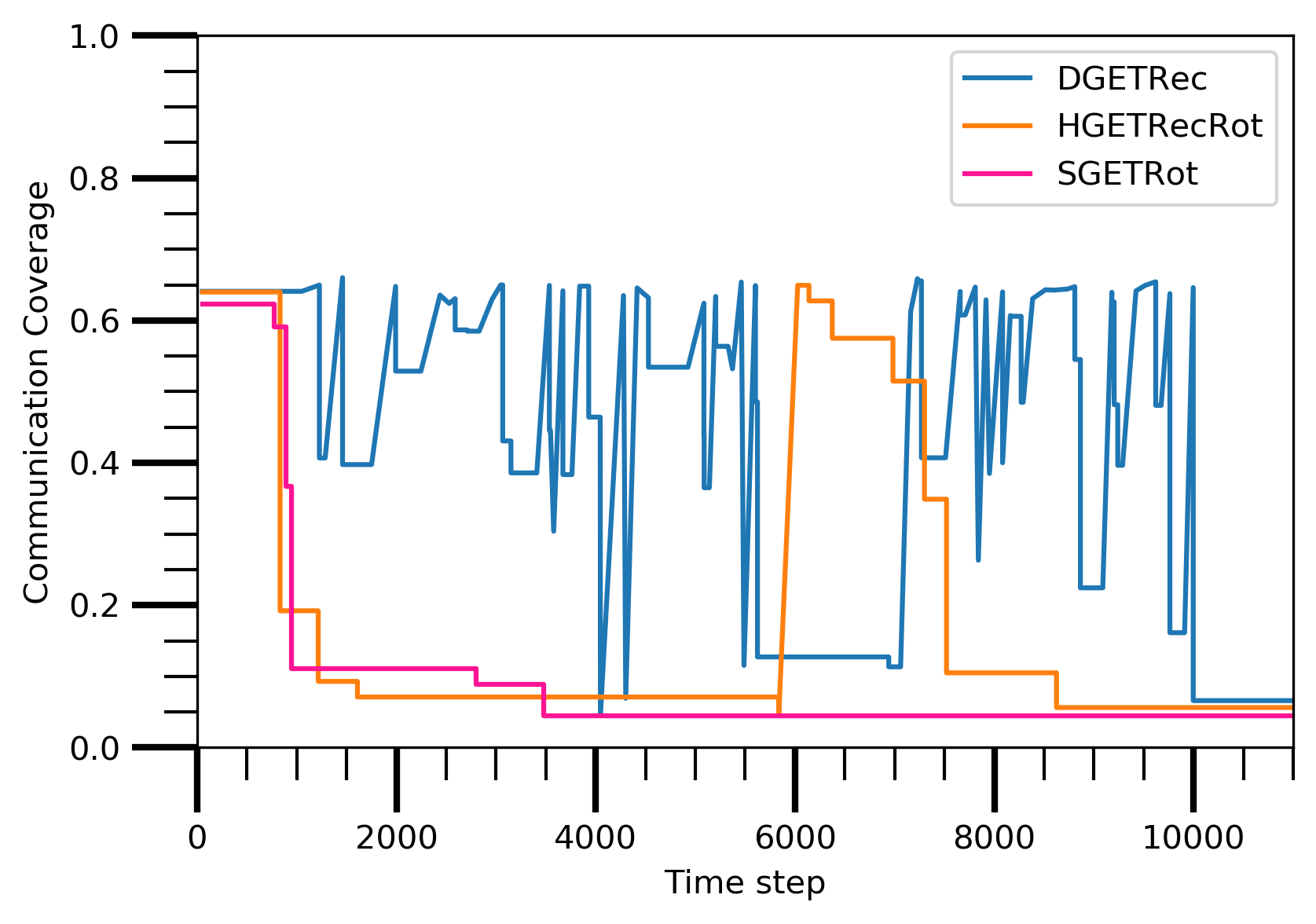}
        \label{A3_energy_Communication_coverage}
    }
    \subfigure[]
    {
        \includegraphics[width=0.45\linewidth]{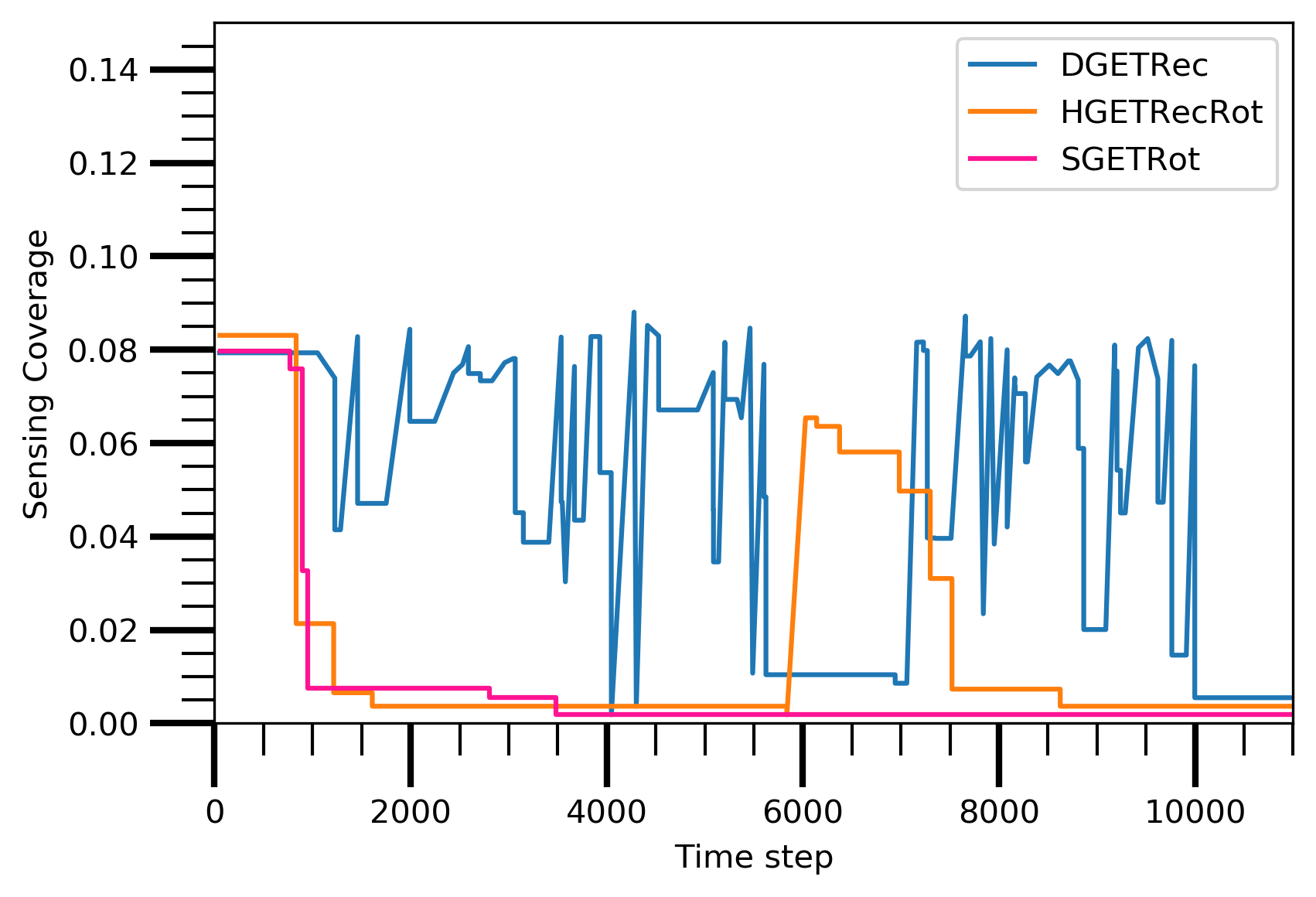}
        \label{A3_energy_Sensing_coverage}
    }
    \caption{Performance of A3 protocol and TM protocols with energy triggering strategy.}
    \label{A3_energy}
\end{figure*}
The TM protocols inclusive of energy triggering methods when used with TC protocols provide the WSN system with a notable outcome. DGETRec, HGETRecRot, and SGETRot and A3 protocols are used for TM and TC, respectively. Furthermore, the performances of the network are shown in figure \ref{A3_energy}. Figure \ref{A3_energy_alive_nodes} represents the number of alive nodes with respect to several time steps. It is seen from the figure that at the initial time steps, DGETRec has the most number alive nodes but at the final time steps, the fact belongs to SGETRot. In figure \ref{A3_energy_Number_of_node_reachable_from_sink}, the number of active nodes reachable from the sink is given in terms of various time steps. The variations in DGETRec, HGETRecRot, and SGETRot are plain observed from that figure. The communication coverage vs. time steps plot is provided, illustrating in figure \ref{A3_energy_Communication_coverage}. It reads the performance of DGETRec in this regard can be addressed better than the HGETRecRot and SGETRot. The affair of HGETRecRot is in between DGETRec and SGETRot. To conclude, the sensing coverage following different time steps is drawn to include in figure \ref{A3_energy_Sensing_coverage}. DGETRec again appears before having enriched sensing coverage than the HGETRecRot and SGETRot. Also, HGETRecRot is not so improved as DGETRec and worse as SGETRot.                      

\subsubsection{A3Cov protocol}
\begin{figure*}[h]
    \centering
    \subfigure[]
    {
        \includegraphics[width=0.45\linewidth]{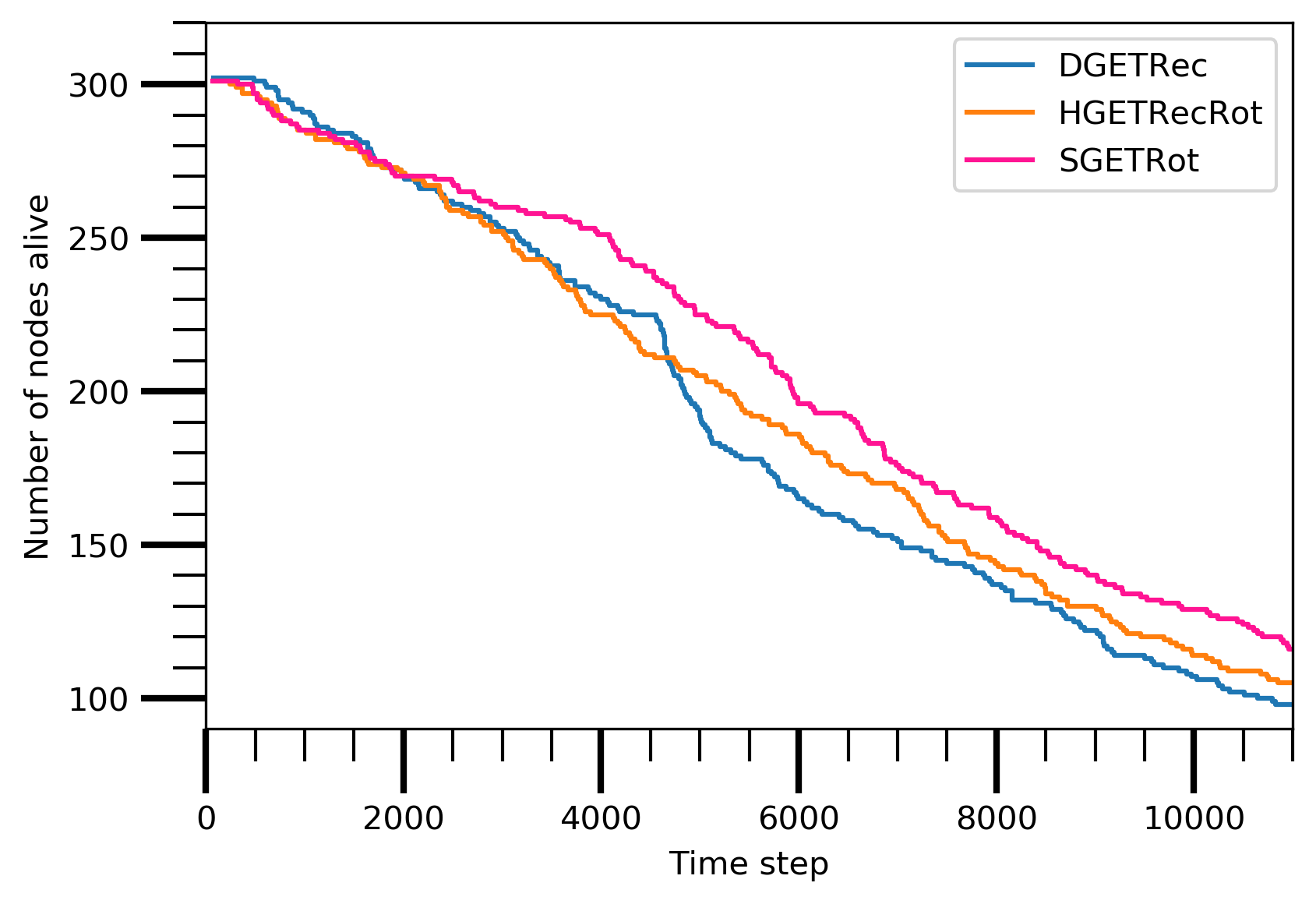}
        \label{A3Coverage_energy_alive_nodes}
    }
    \subfigure[]
    {
        \includegraphics[width=0.45\linewidth]{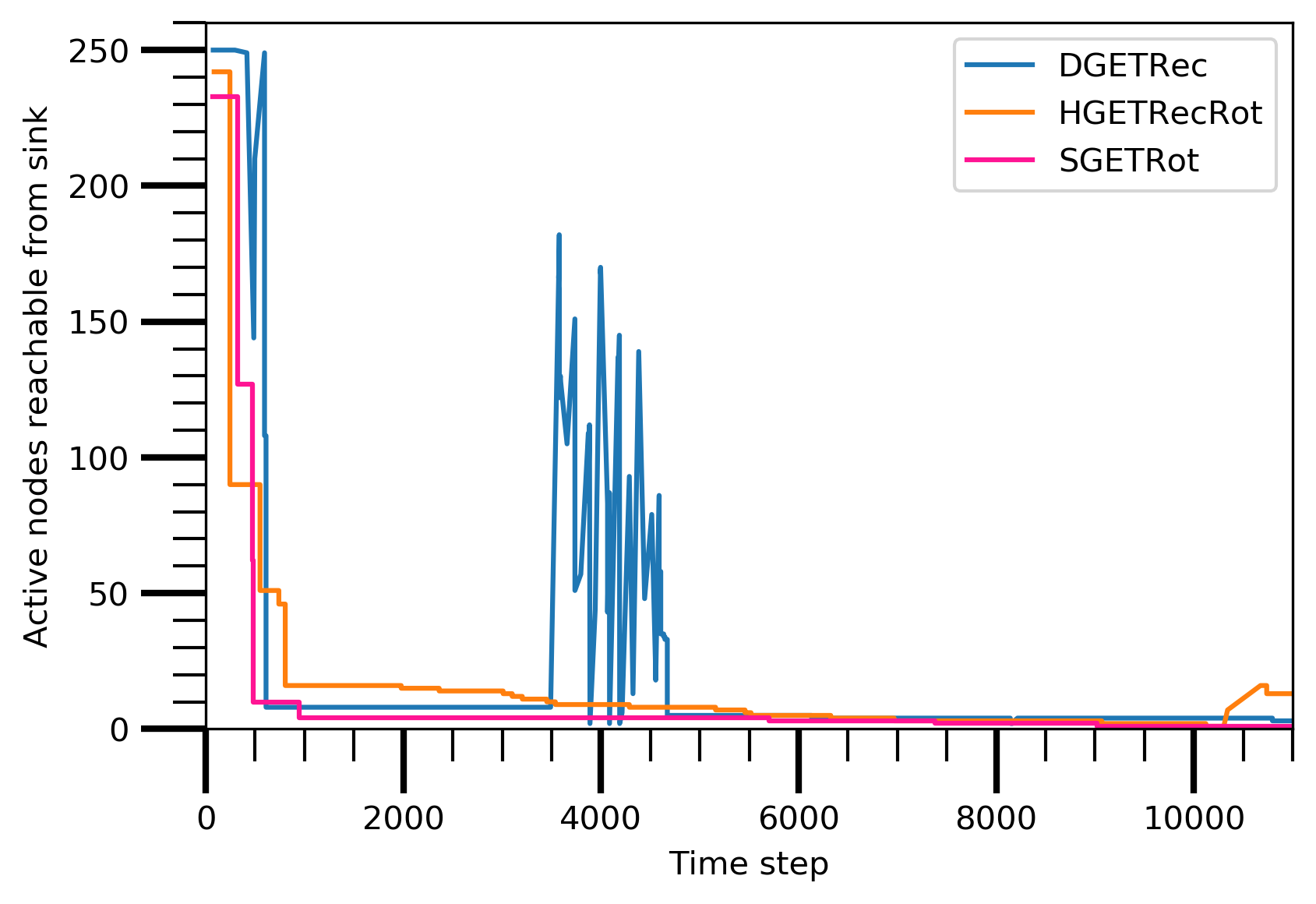}
        \label{A3Coverage_energy_Number_of_node_reachable_from_sink}
    }
    \\
     \subfigure[]
    {
        \includegraphics[width=0.45\linewidth]{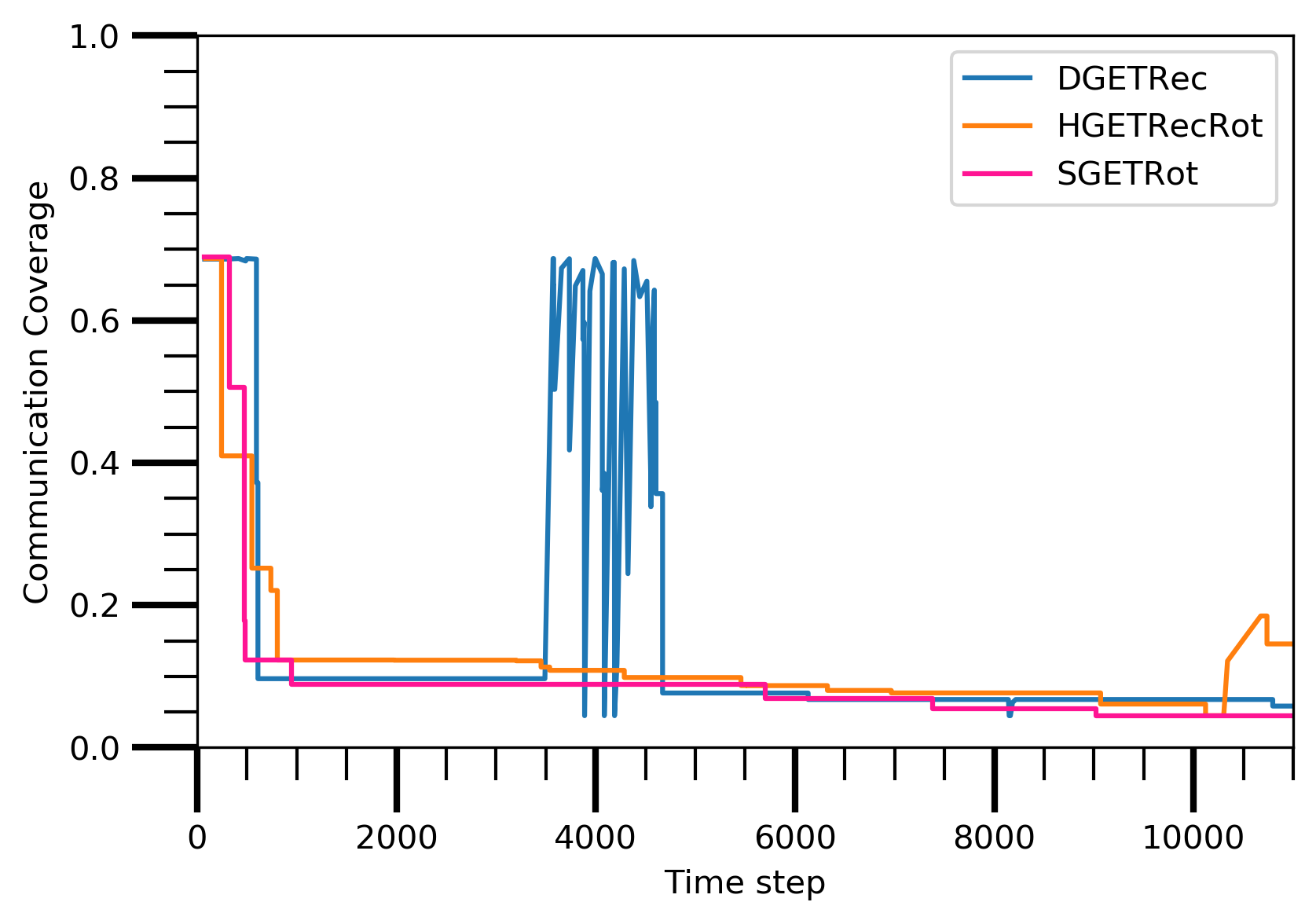}
        \label{A3Coverage_energy_Communication_coverage}
    }
    \subfigure[]
    {
        \includegraphics[width=0.45\linewidth]{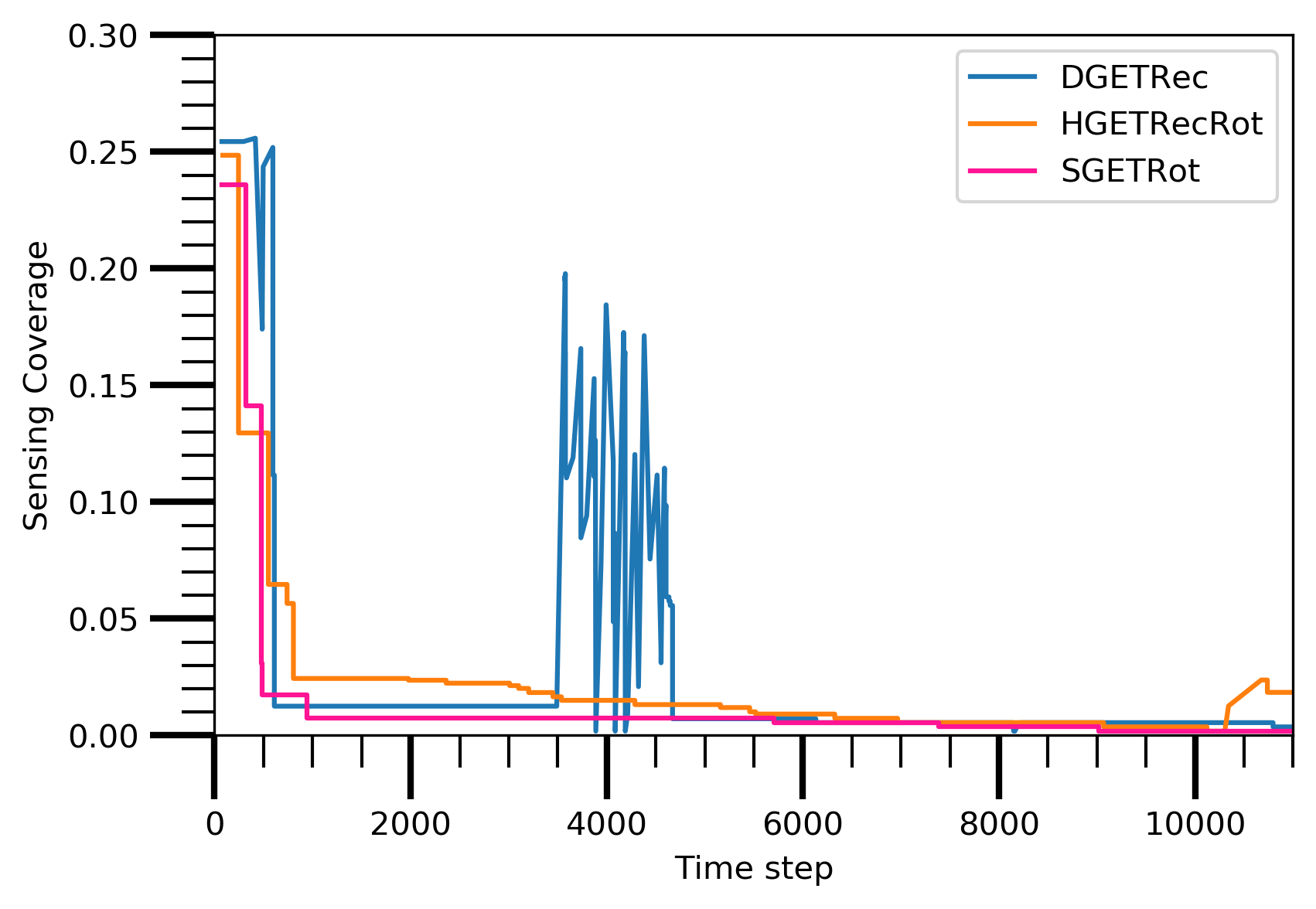}
        \label{A3Coverage_energy_Sensing_coverage}
    }
    \caption{Performance of A3Cov protocol and TM protocols with energy triggering strategy.}
    \label{A3Coverage_energy}
\end{figure*}
In this case, DGETRec, HGETRecRot, and SGETRot TM protocols remain constant but A3Cov replaces the A3 TC protocol. Figure \ref{A3Coverage_energy} provides the network with the ebullition of its performances in this circumstance. The number of alive nodes in pursuit of multiple time steps are delegated in figure \ref{A3Coverage_energy_alive_nodes}. It indicates SGETRot undergoes to have more number alive nodes than HGETRecRot and DGETRec. Additionally, HGETRecRot contains the middle position among the three. A lot of time steps and the number of active nodes reachable from the sink accordingly are arrayed in figure \ref{A3Coverage_energy_Number_of_node_reachable_from_sink}. At the initial time steps and from time steps 3500 to 4700, a few number spikes is occurs in DGETRec. Anything like it is not observed when HGETRecRot and SGETRot are applied similarly. The communication coverage including the consistent time steps are placed in figure \ref{A3Coverage_energy_Communication_coverage}. Some surges are beheld while selecting DGETRec which are absent in HGETRecRot and SGETRot. The sensing coverage with its incidental time steps is demonstrated in figure \ref{A3Coverage_energy_Sensing_coverage}. The figure in fact reflects almost the same pattern produced by the option number of active nodes reachable from the sink. 

\subsection{Time triggering criteria}
\subsubsection{A3 protocol}
\begin{figure*}[h]
    \centering
    \subfigure[]
    {
        \includegraphics[width=0.45\linewidth]{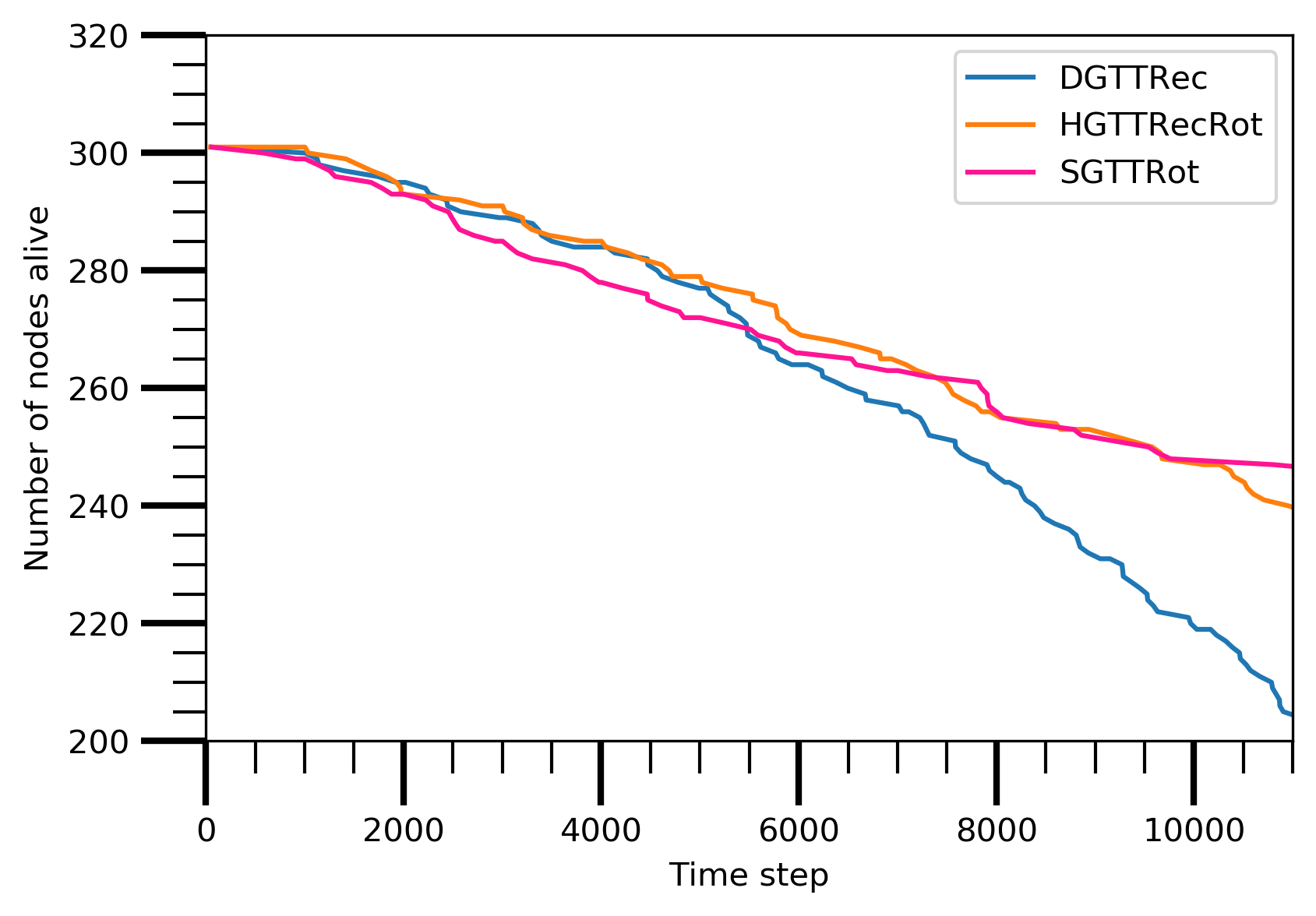}
        \label{A3_time_alive_nodes}
    }
    \subfigure[]
    {
        \includegraphics[width=0.45\linewidth]{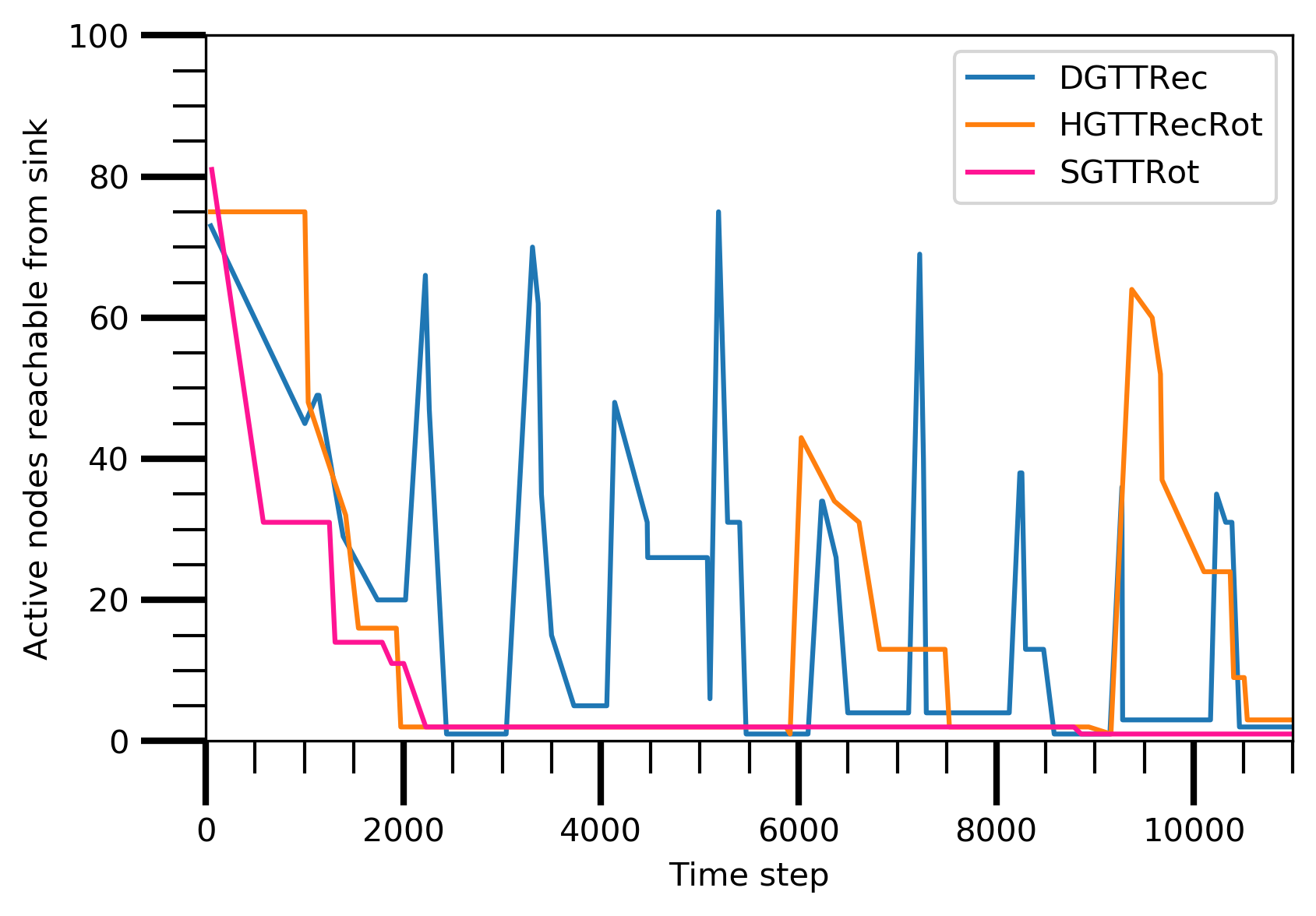}
        \label{A3_time_Number_of_node_reachable_from_sink}
    }
    \\
     \subfigure[]
    {
        \includegraphics[width=0.45\linewidth]{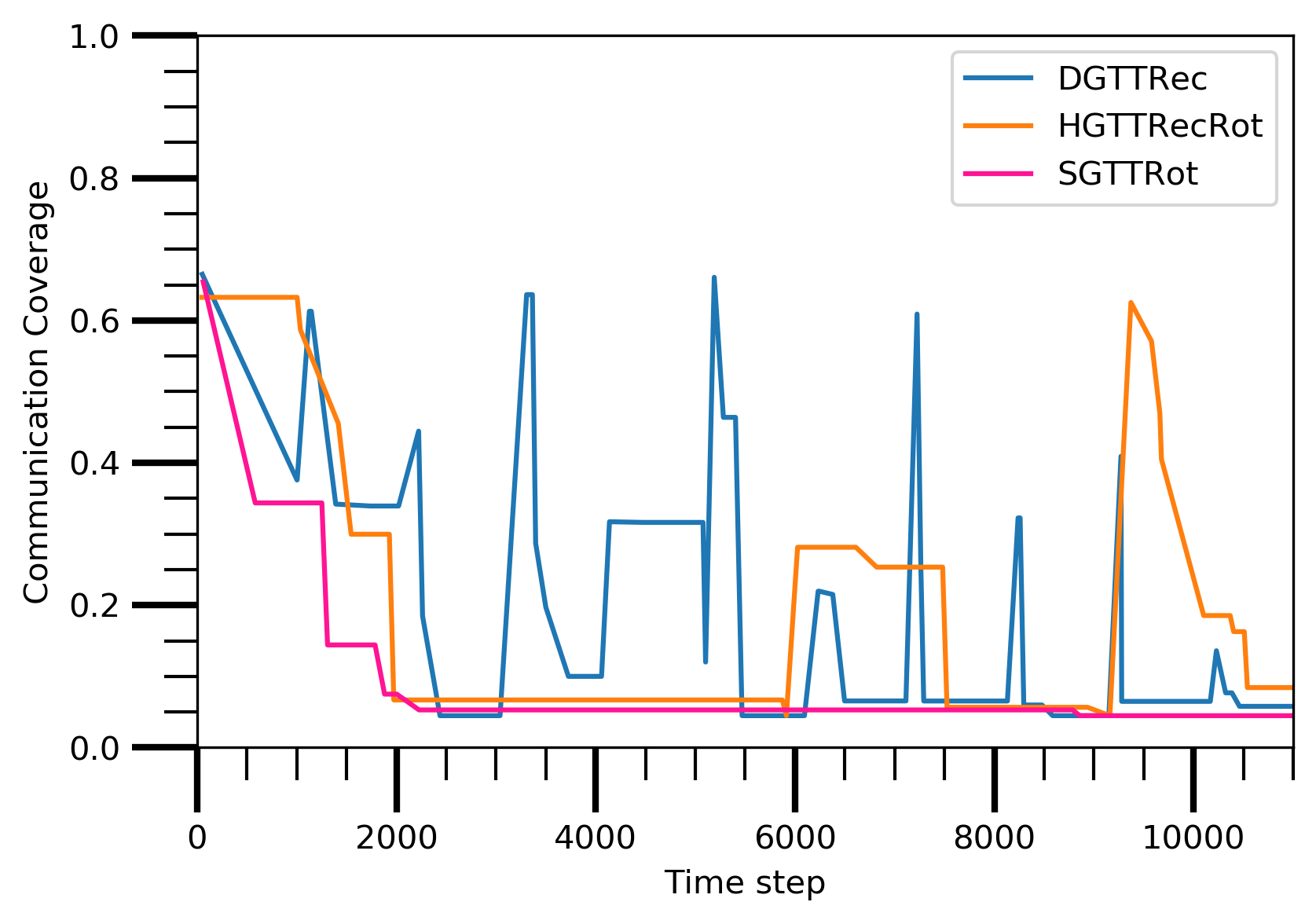}
        \label{A3_time_Communication_coverage}
    }
    \subfigure[]
    {
        \includegraphics[width=0.45\linewidth]{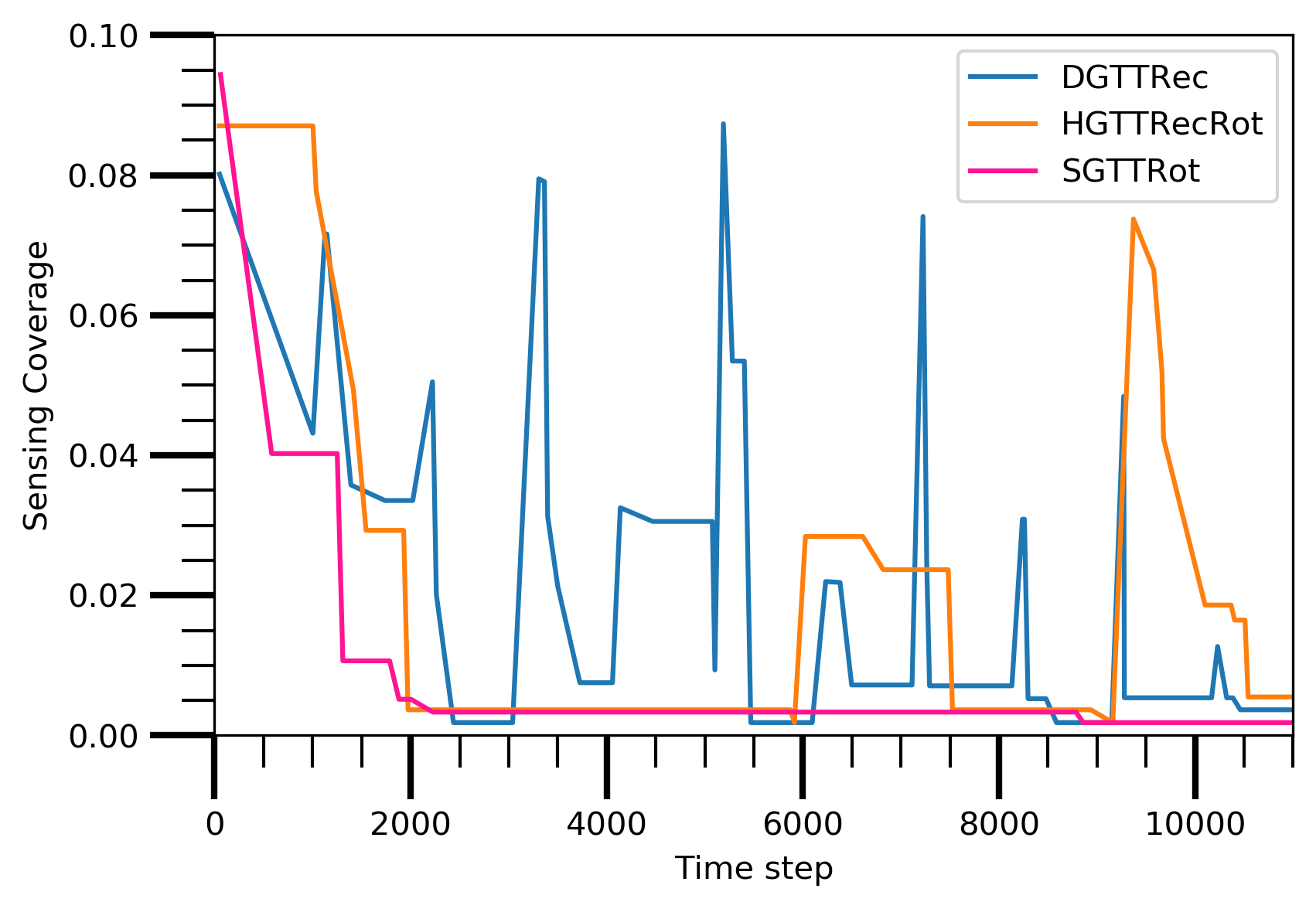}
        \label{A3_time_Sensing_coverage}
    }
    \caption{Performance of A3 protocol and TM protocols with time triggering strategy.}
    \label{A3_time}
\end{figure*}
Time triggering methods integrated TM protocols are unified with TC protocols around the goal of WSN, making the performance noteworthy. DGTTRec, HGTTRecRot, and SGTTRot and A3 protocols are employed of TM and TC, respectively. Moreover, Figure \ref{A3_time} leaves an acquaintance with the performances of the network. Figure \ref{A3_time_alive_nodes} makes familiar the plot involving the number of alive nodes and concurrent time steps. The figure reveals that the performance curve of HGTTRecRot is accompanied by SGTTRot and DGTTRec is far behind them. The number of active nodes reachable from the sink vs time steps is shown in figure \ref{A3_time_Number_of_node_reachable_from_sink}. As the figure, DGTTRec can arrive at the discretion in having the performance flourished than HGTTRecRot and SGTTRot. HGTTRecRot is neither better than DGTTRec and nor worse than SGTTRot. The communication coverage in imitation of the time steps is confined in figure \ref{A3_time_Communication_coverage}. Noted from the figure that DGTTRec has more communication coverage than HGTTRecRot and SGTTRot. The serpentine line of performance in HGTTRecRot is situated in the intermediate place of DGTTRec and SGTTRot. Finally, figure \ref{A3_time_Sensing_coverage} captures the sensing coverage regarding the perspective time steps. It is found in a generous similarity with figure \ref{A3_time_Communication_coverage}.

\subsubsection{A3Cov protocol}
\begin{figure*}[h]
    \centering
    \subfigure[]
    {
        \includegraphics[width=0.45\linewidth]{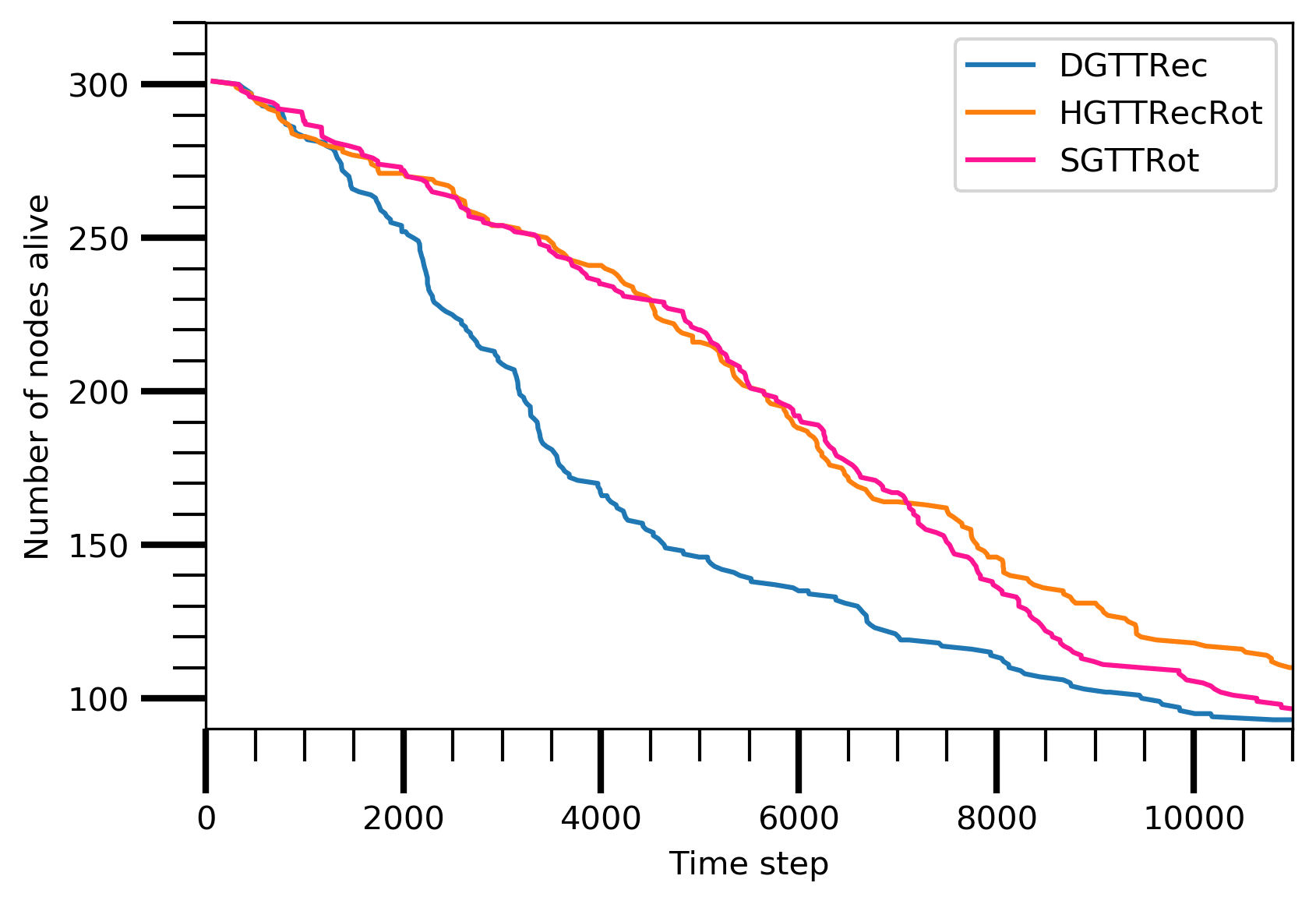}
        \label{A3Coverage_time_alive_nodes}
    }
    \subfigure[]
    {
        \includegraphics[width=0.45\linewidth]{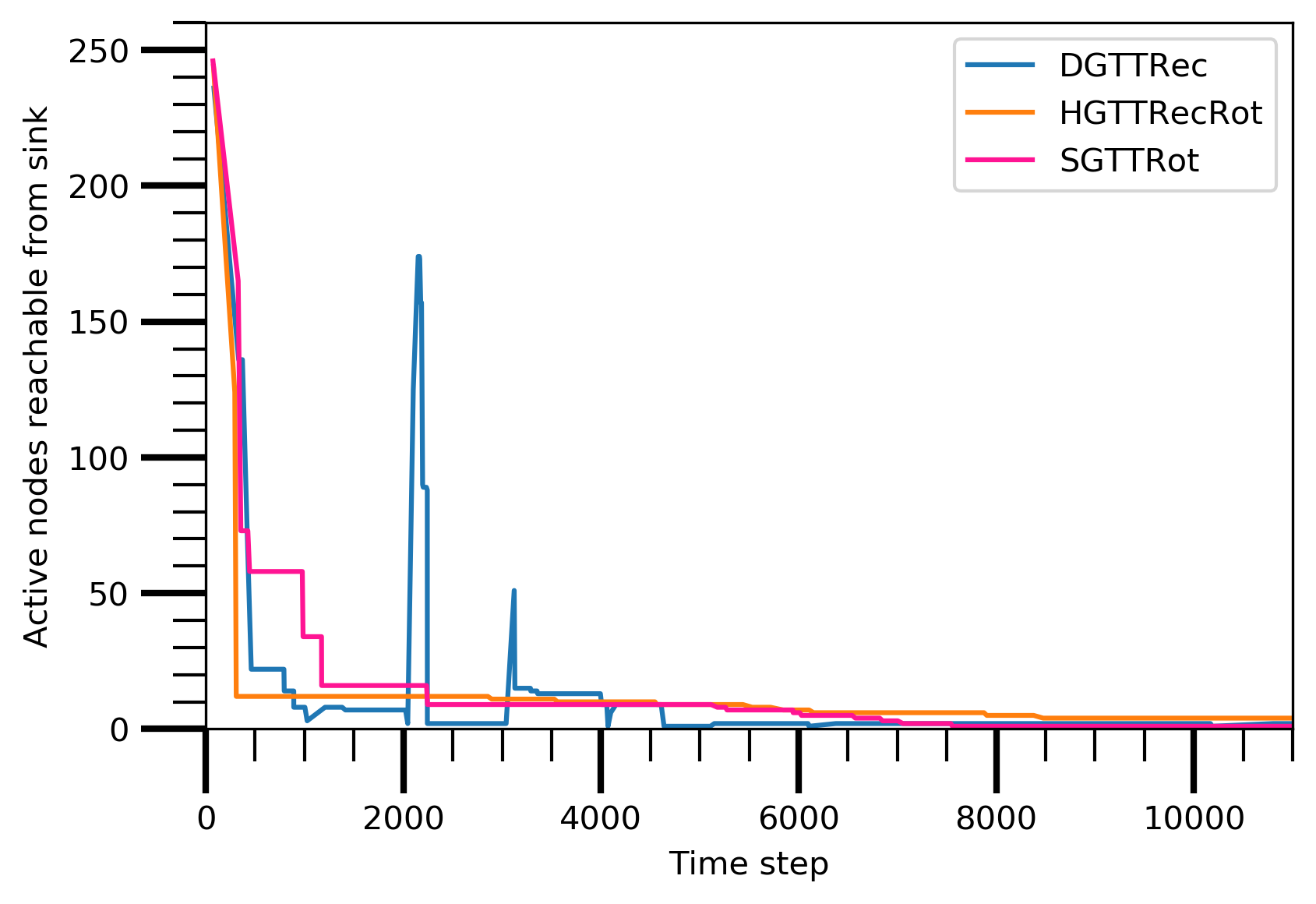}
        \label{A3Coverage_time_Number_of_node_reachable_from_sink}
    }
    \\
     \subfigure[]
    {
        \includegraphics[width=0.45\linewidth]{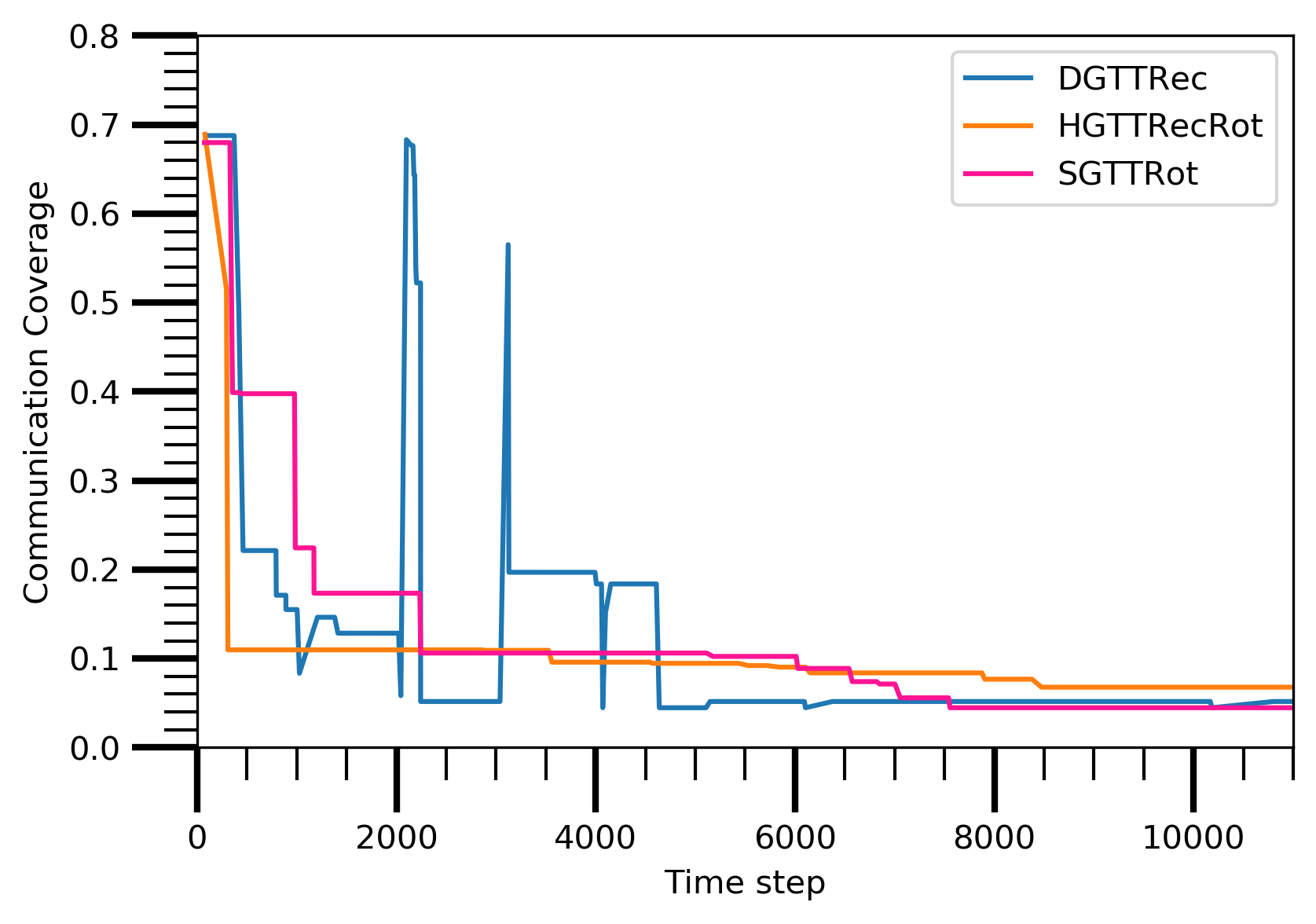}
        \label{A3Coverage_time_Communication_coverage}
    }
    \subfigure[]
    {
        \includegraphics[width=0.45\linewidth]{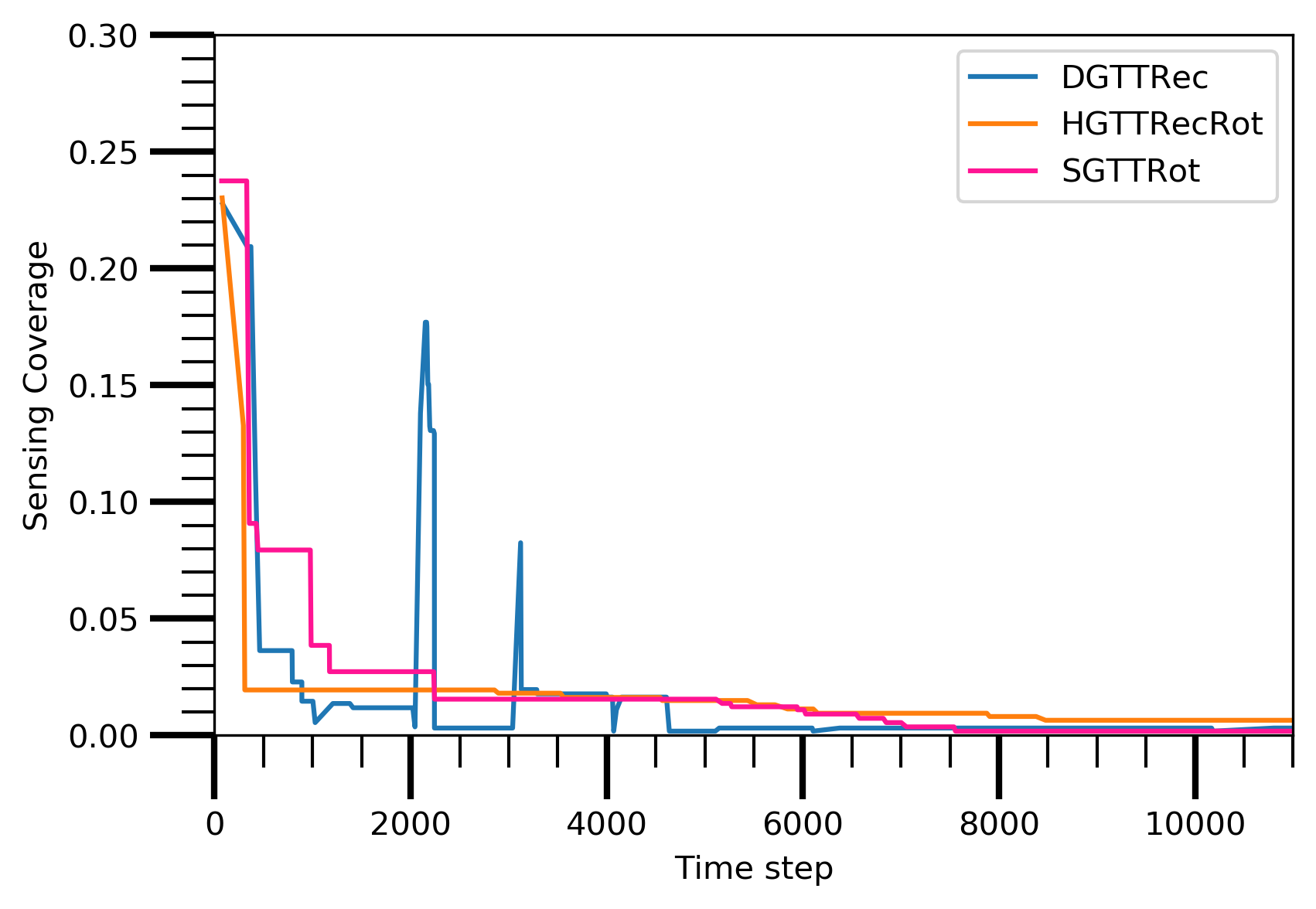}
        \label{A3Coverage_time_Sensing_coverage}
    }
    \caption{Performance of A3Cov protocol and TM protocols with time triggering strategy.}
    \label{A3Coverage_time}
\end{figure*}
Now, A3Cov is substituted in the place of A3 TC protocol, making DGTTRec, HGTTRecRot, and SGTTRot unchanged. On this proviso, the behavior of the network is conserved in figure \ref{A3Coverage_time}. Figure \ref{A3Coverage_time_alive_nodes} is an envoy of the plot consisted of the number of alive nodes after the co-related time steps. It dictates that SGTTRot and HGTTRecRot take to the same course exception of DGTTRec. The number of active nodes reachable from the sink and its associated time steps are marked in figure \ref{A3Coverage_time_Number_of_node_reachable_from_sink}. It is the figure to mark down DGTTRec has a royal end than SGTTRot and HGTTRecRot. Figure \ref{A3Coverage_time_Communication_coverage} narrates the communication coverage on the compatible time steps. A few ripples can be had of in DGTTRec but happening akin to that matter is not viewed from SGTTRot and HGTTRecRot. The sensing coverage in conformity with the time steps are brought to light in figure \ref{A3Coverage_time_Sensing_coverage}. Figure \ref{A3Coverage_time_Sensing_coverage} indeed are come by in the same framework of figure \ref{A3Coverage_time_Number_of_node_reachable_from_sink}.

\section{Conclusion}
\label{conclusion}
This work presents an approach to lengthening the lifetime of WSN-based communication techniques in consideration of an economic zone. An area of 1074m length and 660m width of the economic zone named PowerPac Economic Zone is considered for the deployment of the uniformly distributed sensor nodes. TM protocols are classified into two orders based on the triggering characteristics. DGETRec, HGETRecRot, and SGETRot TM protocols follow the energy-based triggering strategy and DGTTRec, HGTTRecRot, and SGTTRot TM protocols follow the time based triggering strategy. On the other hand, A3 and A3Cov are utilized separately as TC protocol along with all of the TM protocols. Inspections are concentrated basically on two aspects to evaluate the performance of the network lifetime. firstly, the unit combination of TCM protocols and secondly, a relative observation between the triggering strategies of TCM protocols, i.e. time and energy. It is found A3 protocol and TM protocols with both time and energy triggering methods reflects a better performance of the network. Additionally, A3 and DGETRec shapes a long lifespan of the network comparing to the other combinations of TCM protocols. Artificial intelligent networking methods, mobile sensor node deployment, etc. can be incorporated in the future works.

\end{document}